\documentclass[structabstract]{aa}  
\usepackage{natbib}

\usepackage{graphicx}
\usepackage{txfonts}
\begin{document}
   \title{LIDT-DD: A new self-consistent debris disc model including radiation pressure and coupling dynamical and collisional evolution}


   \author{Q. Kral
          \inst{1},
          P. Th\'ebault\inst{1},
          S. Charnoz\inst{2}
          }

   \institute{LESIA-Observatoire de Paris, UPMC Univ. Paris 06, Univ. Paris-Diderot, France
    	\and
	    Laboratoire AIM, Universit\'e Paris Diderot / CEA / CNRS, Institut Universitaire de France
             }

\offprints{Q. Kral} \mail{quentin.kral@obspm.fr}
\date{Received March 4, 2013; accepted August 28, 2013} \titlerunning{The LIDT-DD code : A new self-consistent debris disc model coupling dynamical and collisional evolution.}
\authorrunning{Kral}

 
  \abstract
   {In most current debris disc models, the dynamical and the collisional evolutions are studied separately, with N-body and statistical codes, respectively, because of stringent computational constraints. In particular, incorporating collisional effects into an N-body scheme has proven a very arduous task because of the exponential increase of particles it would imply.}
   {We present here LIDT-DD, the first code able to mix both approaches in a fully self-consistent way. Our aim is for it to be generic enough so as to be applied to any astrophysical cases where we expect dynamics and collisions to be deeply interlocked with one another: planets in discs, violent massive breakups, destabilized planetesimal belts, exozodiacal discs, etc.}
   {The code takes its basic architecture from the LIDT3D algorithm developed by Charnoz et al.(2012) for protoplanetary discs, but has been strongly modified and updated in order to handle the very constraining specificities of debris discs physics: high-velocity fragmenting collisions, 
radiation-pressure affected orbits, absence of gas, etc.
It has a 3D Lagrangian-Eulerian structure, where grains of a given size at a given location in a disc are grouped into "super-particles", whose orbits are evolved with an N-body code and whose mutual collisions are individually tracked and treated using a particle-in-a-box prescription. To cope with the wide range of possible dynamics for same-sized particles at any given location in the disc, tracers are sorted and regrouped into dynamical families depending on their orbits.}
   {The LIDT-DD code has been successfully tested on simplified cases for which robust results have been obtained in past studies: we retrieve the classical features of particle size distributions in unperturbed discs, as well as the outer radial density profiles in $\sim r^{-1.5}$ outside narrow collisionally active rings, and the depletion of small grains in "dynamically cold" discs.
The potential of the new code is illustrated with the test case of the violent breakup of a massive planetesimal within a debris disc. Preliminary results show that, for the first time, we are able to quantify the timescale over which the signature of such massive break-ups can be detected. In addition to the study of such violent transient events, the main potential future applications of the code are planet/disc interactions, and more generally any configurations where dynamics and collisions are expected to be intricately connected.}
   {}


   \maketitle
%

\section{Introduction}
\subsection{Collisional modelling of debris discs}

Debris discs are circumstellar, optically thin dusty discs around main sequence stars with little or no gas\footnote{and should not be mixed up with their younger, gas-rich circumstellar counterparts containing primordial material, known as protoplanetary discs}. Hundreds of such discs have been detected through the IR-excess associated to the warm and/or cold circumstellar dust in the micron to centimetre size range (see reviews by \citet{wyat08} and \citet{kriv10}).

The observed dust cannot be primordial, because its removal time scale due to Poynting-Robertson drag and destructive collisions is much shorter than the system's age. It must thus be steadily replenished from a mass reservoir contained in larger, unseen bodies, most probably through a collisional cascade starting at bodies exceeding the kilometre-size range. Studying the collisional evolution of such systems is thus of fundamental importance, as it allows to address crucial issues such as the size-distribution of particles in the disc, the link between the observed dust and its hidden larger progenitors, the total mass in the system and its mass loss with time. All these issues are usually numerically investigated with particle-in-a-box codes, with no or poor spatial resolution, dividing the disc into size bins whose mutual collisional interactions are treated in a statistical way \citep{keny02, theb03, kriv06, theb07, lohn08, gasp12}. Such models have greatly improved our understanding of debris discs in the past decade and are essential to understand the formation and evolution of planetary systems.

\subsection{Dynamical modelling}

For a minority of debris discs (40 to date\footnote{see http://www.circumstellardisks.org}), resolved images have also been obtained. These images are usually obtained at visible and near-IR wavelengths, but new observational facilities such as the Herschel Space Observatory are now also providing more and more images in the mid to far IR \citep{lohn12,wyat12,lest12,dona12,erte12b}. Almost all of these resolved images show pronounced structures, such as bright clumps, two-side asymmetries, spiral arms or warps \citep[e.g.][]{kala05,goli06,acke12,boot13}. Understanding the origin of these structures has been a major objective of debris disc studies, which have explored, mostly through numerical modelling,  several possible scenarios. In many cases, these scenarios involve the presence of a perturbing planet, such as for the warp of the $\beta$ Pictoris disc \citep{moui97,auge01}, the brightness asymmetries in the Epsilon Eridani system \citep{kuch03} or the confined rings around HR4796A and Fomalhaut \citep{wyat99,chia09}. However, alternative scenarios, such as stellar perturbations \citep[e.g.,][]{auge04,theb10}, interaction with gas \citep{take01,besl07}, interaction with the ISM \citep{arty97,debe09,marz11} or transient violent events \citep{keny05, grig07} have also been investigated.

The numerical exploration of these different scenarios has mostly been done using $N$-body codes, which are designed to accurately follow the development of dynamical structures such as resonances, migrations, etc \citep[e.g.,][]{rech09, kuch03}. The price to pay for the spatial precision allowed by the N-body approach is that collisions are usually neglected in such models, which follow the evolution of collisionless test particles.

\subsection{The need for coupled studies}

As can be seen, the collisional and dynamical evolution of debris discs are usually studied separately, the focus of collisional models being mostly global characteristics such as grain size distributions or mass loss, while dynamical models focus on local or transient phenomenae that can leave signatures in resolved images. While such separate studies can (and have) produce(d) important results, they suffer from unavoidable limitations.

For the purely statistical models these limitations are obviously the absence of or poor spatial resolution, but also the fact that dynamical processes might strongly affect impact rates and velocities and thus the collisional evolution. This might lead to strong discrepancies in the way some parts of the disc collisionally evolve with respect to others.

The N-body collisionless models also suffer from severe handicaps. The absence of collisions can indeed strongly bias or even invalidate results obtained in such codes, and this for several reasons. As an example, if collisional timescales are shorter than dynamical ones then collisions can hinder or even prevent the build-up of dynamical structures. Likewise, the identification of dynamically stable and unstable regions in a perturbed debris disc (be it by a planet or a stellar companion) can also be strongly affected by collisional activity, as a collision cascade will steadily produce small grains that can be launched by radiation pressure on highly eccentric or unbound orbits that can populate regions that are in principle dynamically "forbidden". 

Unfortunately, the complex interplay between the system's dynamical and collisional evolutions is very difficult to handle numerically, mainly because the particle-in-a-box and the N-body approaches are radically different in their principles and structures. However, the first attempts to partially couple dynamics and collisions have been published recently, most of them taking as a basis the N-body approach into which some collision-imposed properties are injected.
 
\subsection{First attempts at coupling dynamics and collisions}\label{firstatt}
\noindent

The most reliable way of including destructive collisions into an $N$-body scheme would in principle be by "brute-force" methods, where bodies are effectively broken into fragments whose evolution is then dynamically followed \citep{beau90}. However, such codes can only follow a very limited number of collisional fragments and lead in any case to an exponential increase of the number of particles that very quickly becomes unmanageable. 

An alternative option is to run collisionless $N$-body runs, and to post-process them assuming that each test particle stands for a dust-producing collisional cascade of solids \citep{boot09}. This approach allows to explore very long timescales, up to several Gyrs. It has been used to study the signature, in terms of Spectral Energy Distribution (SED), of the late stages of terrestrial planet formation \citep{raym11}. However, it has some important limitations, as it implicitly supposes that collisions and dynamics are fully decoupled, i.e., assuming that collision processes are unaffected by the dynamics, but also that collisions have no influence on the formation and fate of spatial structures.

An important recent improvement is the LIPAD code developed by \citet{levi12}. LIPAD follows the dynamical evolution of "tracers", representing populations of single-sized planetesimals and tracks down their mutual collisions. Collisions change tracer velocities as well as increase or decrease the planetesimal sizes they represent depending on the collision outcome (accretion or fragmentation). However, this code is designed to study the earlier stage of kilometre-sized planetesimal accretion and is not adapted, in its present form, to debris disc studies. It indeed focuses mainly on the fate of the large growing bodies and has a very simplified prescription for the dust: all material below the minimum planetesimal size is into "dust tracers" having a single physical size and that no longer collisionally interact with each other. Furthermore, it neglects the effect of radiation pressure, an effect that is crucial for debris discs because it creates a strongly size-dependent behaviour of dust particles, and thus greatly complexifies the numerical treatment of the dust population.

Amongst the most sophisticated debris discs' models that have been developed to date are probably the "CGA" \citep{star09} and "DyCoSS" \citep{theb12a} algorithms. These codes are different in their principles but both implement comparable levels of collisional processes into their $N$-body main structure. They are specifically designed to estimate how collisional lifetimes of grains are affected by dynamical perturbations, and how this variety of collisional lifetimes in turn affects the development of dynamical structures (resonances, PR-drag migrations, etc...). Both codes allow the study of very fine spatial structures and have been used for some specific cases such as the Neptune-Kuiper-Belt system \citep{kuch10}, debris discs in binaries \citep{theb12a} or discs with an embedded or exterior planet \citep{theb12b}. They are however both restricted to specific set-ups, i.e., systems at collisional and dynamical \emph{steady-state} under the influence of \emph{one} perturbing body only. Moreover, the coupling between dynamics and collisions is only partial and both codes assume that collisions are fully destructive, i.e., particles having exceeded their collisional lifetimes are simply removed, and the fate of small collisional fragments is not followed.

\subsection{"Tracer" approach}\label{spapproach}

The first attempt at truly coupling the collisional and dynamical modelling of a debris disc was performed with the hybrid code of \citet{grig07}. The principle of this approach is that each particle of an $N$-body simulation is a "tracer" (or a "super-particle") representing a vast population of particles of a given size, and that the code tracks mutual collisions between these tracers. These collisions are then treated with a classical particle-in-a-box approach, estimating the amount of mass lost by the impacting tracers and the corresponding mass that is injected into new tracers carrying the collisional fragments created. This code was successfully used to identify and quantify the mechanism of collisional "avalanches" produced by the shattering of a large planetesimal far inside a dense dust belt. It had however one major limitation, which is that the number of tracers is constantly increasing and rapidly becomes too much for the code to handle. As such it was restricted to short timescales, typically a few orbital periods.

This major limitation has been recently overcome by an algorithm aimed at studying the very different astrophysical case of young, massive and gaseous protoplanetary discs \citep{char12}. This code, named LIDT3D, is based on a similar "tracer" approach but integrates a routine to identify superfluous and redundant tracers and to reassign these tracers to regions where they are more needed. This procedure prevents the number of tracers from increasing and allows studies spanning much longer times. However, LIDT3D cannot be used to study debris discs, because of the very different physical processes at play in a protoplanetary disc, in particular the strong homogenizing effect of gas drag that greatly simplifies the dynamics, as well as the very different collisional regime that prevails, i.e., mostly low-velocity impacts instead of high-$\Delta v$ shattering-and-fragment-producing ones in debris discs.

Our aim is here to create a new version of the LIDT code, called "LIDT-DD", which keeps the basic tracer architecture of LIDT3D, but is able to handle the very constraining demands of modelling debris discs. We present in Section 2 a brief summary of the main LIDT architecture that will be kept in the new algorithm. In section 3, we first describe the challenging aspects of debris disc physics and their implications in terms of numerical handling. We then present the LIDT-DD code itself as well as an illustrative pedagogical run. Section 4 is devoted to different tests that have been performed to check the code's reliability. In Section 5, we illustrate the potential of this new code by showing some preliminary results for an example set-up of a massive catastrophic collision within a debris disc.  


\section{LIDT basic architecture}

The basic architecture of the LIDT-DD debris disc model takes its roots in the LIDT3D code (OPEN-MP) developed by \citet{char12} for the study of primordial protoplanetary discs. Because of the radical differences between these two astrophysical cases, in terms of dominant physical mechanisms but also regarding the strong constraints imposed by high-velocity destructive collisions in debris discs (see Sec. \ref{specdd}), LIDT-DD strongly departs from its predecessor, incorporating several new key procedures, and can be considered as an independent, stand-alone code. However, before presenting it in great details in the next section, let us first briefly recall here the basic features of the LIDT code that remains as the backbone of this new model \citep[for a more detailed description, see][]{char12}.

\subsection{Tracers}

The building blocks of the LIDT scheme are "tracers", each representing a whole population of particles of a given size at a given location in the system. At each time step, these tracers are dynamically evolved with a deterministic N-body integrator and do collisionally interact following a statistical particle-in-a-box scheme.

Tracers are defined by their position $\vec{R_\textrm{\tiny{t}}}$, velocity $\vec{V_\textrm{\tiny{t}}}$, the physical size $s_\textrm{\tiny{t}}$ of the particles they stand for and the individual mass $m_\textrm{\tiny{t}}=4/3 \pi \rho s_\textrm{\tiny{t}}^3$ of these particles. A tracer represents a large number $N_\textrm{\tiny{t}}$ of such individual particles, and thus represents a total mass $M_\textrm{\tiny{t}}$ such as $M_\textrm{\tiny{t}}=N_\textrm{\tiny{t}} \, m_\textrm{\tiny{t}}$. The whole system is then composed of $N$ such tracers, a number that can evolve with time depending on the number of tracers that are needed in every region of the disc. For our LIDT-DD debris disc simulations, $N$ is of the order of $\sim 10^5$, which is the typical maximum total number of tracers that are needed in our simulations to handle 2 tracers per dynamical category per size bin per spatial cell per time step (see Sect.~\ref{sorting}). $N$ will be worked out automatically by the code at every time step as the number of tracers directly depends on the dynamical and collisional activity within the disc (see section \ref{reassignment} for more details).

\subsection{Dynamical Evolution}

The dynamical evolution of a tracer representing a cloud of particles of size $s$ is followed using a lagrangian N-body approach, integrating at each time step the equation of motion corresponding to one particle of size $s$ at position $\vec{R_\textrm{\tiny{t}}}$ with velocity $\vec{V_\textrm{\tiny{t}}}$. Forces that are taken into account are the central star's gravity, stellar radiation pressure and friction due to turbulent gas. The integrator is a Burlisch-Stoer with a semi-implicit solver \citep{bade83}.

\subsection{Collision treatment} \label{collt}

Once all tracers have been dynamically evolved for a time step, their collisional evolution is computed in the following way: the system is divided into a 2-D grid, in $(r,z)$ or $(r,\theta)$, of spatial cells, into which the collisional evolution will be followed. Within each cell, mutual collisions between tracers of each size group are considered. The first step is to compute average impact velocities $\Delta V_{i,j}$ between all pairs of sizes $s_i$ and $s_j$. Then, from the values of $\Delta V_{i,j}$ and the particles number $N_i$ and $N_j$, the number of collisions $Nc_{i,j}$ between all particles contained in tracer i and those contained in tracer j within a time step $\Delta t$ is derived, following:

\begin{equation}
	Nc_{i,j} \sim \frac{\Delta V_{i,j} . N_i . N_j . \sigma_{i,j} }{\mathrm{Vol}_{i,j}} \Delta t,
 	\label{eqCr}
\end{equation}

where $\mathrm{Vol}_{i,j}$ is the cell's volume and $\sigma_{i,j}=\pi(s_i+s_j)^2 $ is the total cross section during the impact.

Then, for each impacting pair ($i,j$) a standard collisional outcome prescription is plugged in to estimate if the collision results in accretion, rebound or erosion. In the latter case, the number of collisional fragments produced for each size is estimated using simple collisional laws \citep[see][]{char12}. For each fragment size, the number of produced fragments is then added to the tracers present in the cell corresponding to this size. If no corresponding tracers are already present, a new one is created. 

A key feature of the LIDT scheme is a tracer reassignment procedure that avoids a unmanageable increase of the number of tracers. Its basic principle is that each redundant tracer, when for instance there are too many tracers of a given size in a given cell than would be required to give a statistically significant behaviour, is taken away (its mass being given to other neighbouring tracers of the same size) and stored in a buffer of "freed" tracers that can be used whenever new tracers are created by collisions in other cells.

To avoid too fast movements of the tracers in and out of the cells, which would put an artificial constraint on the time step, the $(r,\theta)$ grid is rotating at the Keplerian velocity calculated at the center of each cell. 

Note that the collisional spatial grid has necessarily a finite spatial extension and cannot extend over the whole space where the dynamical evolution of tracers is followed. There is thus a region beyond the outer limit of the grid where collisions are implicitly not taken into account, but this is an acceptable simplification if the outer limit of the grid is located in regions that are sparsely populated and not very collisionally active.

As a summary, the basic principle of the LIDT scheme can be schematically presented as follows:

\begin{itemize}
      \item \it{Step 1}: Evolve dynamics,
      \item \it{Step 2}: Create a 2-D grid superimposed onto the tracers to divide the system in different spatial cells,
      \item \it{Step 3}: Compute relative velocities between tracers in each cell,
      \item \it{Step 4}: Compute collisional outcomes and produce collisional fragments in each cell,
      \item \it{Step 5}: Create tracers and reorganize them within the system.
\end{itemize}

\section{LIDT-DD}\label{lidtdd}  

\subsection{Specificities of debris discs physics}\label{specdd}

The major changes that have been implemented into LIDT-DD with respect to the initial protoplanetary-disc version have been motivated by the specificities of debris discs physics, which impose very strong constraints on the way collisions and dynamics are to be treated. We present here a brief description of what these specificities are and how they will affect the numerical treatment of the system within the LIDT frame.

One crucial difference between protoplanetary and debris discs is the absence (or strong depletion) of gas in the latter. While this absence is a simplification in terms of the physical processes at play, in particular allowing to dispense with the problematic handling of turbulence and gas disc parameterizing \citep[see][]{char12}, it does in fact add great complexity to the system's dynamical and collisional evolution. Indeed, in a protoplanetary disc, small grains are strongly coupled to the gas, and gas drag very quickly smoothes out any disparities between orbits of similarly sized particles. As a consequence, the way these particles are produced and the orbits on which they are released is of little importance, as any initial conditions are very quickly relaxed. The welcome consequence is that, at any given location in the disc, there is only \emph{one} possible dynamics for a given particle size. In numerical terms, this means that, in a given "cell" of the LIDT code, only one category of tracer is needed to represent one size bin. 
This is unfortunately no longer the case for debris discs, where the absence of gas drag makes that \emph{initial conditions are never relaxed}. So that, depending on where and how they are produced, particles of the same size can have very different dynamics despite being in the same region of the system. This puts very strong constraints on the numerical treatment, as several categories of tracers might be needed for each size bin in each cell. A collateral problem is that this number of categories cannot be known in advance. Another consequence is that the way new particles (or their tracers) are produced suddenly becomes an issue, as this will control their fate and the fate of the new particles they will in turn spawn because of later collisions; whereas in LIDT3D tracers are simply automatically given the only possible gas-imposed dynamics in the region they are produced. 

Another major difference with the protoplanetary disc case is that impact velocities are much higher, partly because the damping effect of gas drag is no longer present, but also because debris discs are expected to be dynamically stirred by massive (and often unseen) bodies \citep[see, e.g.,][for a discussion on this issue]{theb09}. These high velocities have dramatic consequences on the treatment of collisions, because they will lead to very destructive impacts producing numerous small fragments of all sizes. In the protoplanetary disc case, no such high-dv impacts are expected, so that the treatment of collisions was handled in a very simplified way, i.e., by setting a threshold velocity of 1\,m/s, regardless of the impacting bodies sizes and compositions, beyond which all impacts were considered as "erosive", and all erosive impacts resulted in similar outcomes in terms of mass loss and fragment distribution. Such simplified laws cannot be used for debris discs, because the outcome of fragmenting impacts strongly depends on parameters such as colliding bodies mass ratio and velocities \citep[e.g.,][]{benz99}. The consequences of, say, a 100\,m/s impact and a 1\,km/s one are radically different in terms of mass loss, size of the largest remaining fragment or size distribution of the produced debris. 

The last fundamental specificity of debris discs is the crucial role played by grains very close to the blowout size $s_\mathrm{cutoff}$ imposed by radiation-pressure. Using the parameterization with the ratio $\beta=F_\mathrm{PR}/F_\mathrm{grav}$, which is $\propto 1/s$ in a wide size range, then these grains correspond to values very close to $\beta=0.5$ \footnote{for the simplified case where they are produced from parent bodies on circular orbits}.
As has been shown by most collisional evolution models, the total geometrical cross section in debris discs, and thus their luminosity at all wavelengths short of the mid-to-far IR, is dominated by solids in the $s_\mathrm{cutoff}$ to $\sim 2-3 \, s_\mathrm{cutoff}$ range \citep[e.g.][]{theb07}. Unfortunately, contrary to protoplanetary discs where gas coupling makes them behave very similarly to grains of other sizes, in gas-poor discs these small grains have a very complex dynamical and collisional evolution. They are indeed placed by radiation pressure on high-eccentricity orbits, making them populate vast regions extending very far from their production location. Conversely, small grains present at any given location in the disc can potentially originate from far away regions much closer to star. Furthermore, small grains have strongly varying orbital (and thus impact) velocities along their elongated orbits. Last but not least, close to the $s\sim s_{cutoff}$ limit, this complex dynamical behaviour becomes extremely sensitive to very small size differences. Indeed, a grain with a high $\beta$ produced from a circular orbit of semi-major axis $a$ has an apoastron
\begin{equation}
	Q= a\,\,\frac{1-\beta}{1-2\beta} \, \left(1\,+\,\frac{\beta}{1-\beta}\right)=\frac{a}{1-2\beta},
 	\label{apobeta}
\end{equation}
So that grains with $\beta=0.48$ have an apoastron of $25a$, while $\beta=0.45$ grains, which are just $7\%$ bigger, reach only $10a$.
These extreme characteristics of the particles that happen to be the ones dominating disc luminosities put very strong constraints on the way the code has to handle the critical $\sim [s_\mathrm{cutoff},2s_\mathrm{cutoff}]$ size range.

\subsection{Dynamical evolution}

The dynamical computation part of the code is very modular and could be adapted to the present case. 
One important update made to fit debris disc physics was to add Poynting-Robertson drag, following the equation \citep{robe37}
\begin{equation}
\frac{d^2 \vec{r}}{dt^2}= -\frac{GM_*}{r^2}\, \vec{e_r} + \frac{\beta GM_*}{r^2} \left[ \left( 1 - \frac{\vec{v}.\vec{e_r}}{c} \right) \vec{e_r} - \frac{\vec{v}}{c} \right],
  \label{PRdrag}
\end{equation}
where $\vec{r}$ is the position vector for a particle in the frame of its central star (of mass $M_*$), r is its norm, $\vec{v}$ is the velocity vector, $c$ is the speed of light and \vec{e_r} is the radial normalized vector used in spherical coordinates. Note, however, that PR-drag is never dominant in the highly collisional regime that is considered throughout this paper. Another important change was to take into account the possibility to handle dynamical perturbing bodies such as planets or stellar companions.

As in LIDT3D, the dynamical evolution of the tracers is followed beyond the outer limit of the collisional grid (see Section \ref{collt}). However, to avoid computing the evolution of irrelevant tracers too far from the region of interest, an outer limit $r_{outDyn}$ for following their orbits is also imposed, beyond which tracers are considered as lost and are removed from the system. Contrary to the protoplanetary disc case, this external border has to be located relatively far out in order to follow the orbits of the population of small, high-$\beta$ particles that can have their apoastron very far from the central star. Within the rest of the paper the dynamical outer border has been fixed to $r=300$ AU so that the smallest bound particles taken into account in the simulations can come back and collide into their birth ring. 

\subsection{Particle size sampling}\label{sampling}

As has been seen in sec. \ref{specdd}, small high-$\beta$ grains are of extreme importance as they dominate the disc's luminosity for most observations. Furthermore, we have seen that these grains' behaviour is very sensitive to small size differences. As a consequence, the resolution between adjacent size bins must be small enough to have a correct sample of grain sizes in this high-$\beta$ domain. 

Another major constraint on the size distribution is that it should stretch up to bodies that are large enough to sustain collisional cascades on long timescales, i.e., bodies of size $s_{max}$ should have a collisional lifetime exceeding the simulations' timescale. This constraint will vary depending on the disc's stirring, particle composition and of course on the considered timescales. To be on the safe side, we take $s_{max}=50$\,km, which corresponds to the smallest primordial bodies found by \citet{lohn08} for their highly-stirred simulation after $5\times 10^{9}$years.

With a classical size sampling where size bins are separated by a constant increment in logarithmic scale, these two constraints put together would lead to a number $N_b$ of size bins that is much too large to be handled in the simulations, for which this number can typically not exceed $\sim60$. However, these constraints can be relaxed, because a very fine size-sampling is only needed in the small-size range but not for larger bodies, for which dynamical behaviours no longer have extreme variations with size.
For these reasons we consider two different scales, a fine sampling for high-$\beta$ grains, and a coarser one for all the other sizes. For the fine sampling domain, the logarithmic increment in size $\epsilon_f$ is taken as a free parameter in the [1.05,1.25] range. This choice of $\epsilon_f$, combined with the value for $N_b$, then automatically constrains both the limiting size $s_{lim}$ between the fine and coarse domains, as well as the size increment $\epsilon_c$ in the latter, which is always within the [1.8,2.1] range. Note that the "coarse" size increment does not reach too high values, so that the code does not loose too much precision in the size-dependent treatment of collisions. As a matter of fact, it is comparable to the size increment considered for the ACE code \citep{kriv06}. Within the rest of the paper we consider a nominal case with $\epsilon_f = 1.15$ for the smallest size bins and $\epsilon_c=2$ for the coarse domain.

\subsection{Impact velocity estimates and collision search}\label{vcol}

For the case of protoplanetary discs, impact velocities $\Delta v_{i,j}$ in a given cell were simply derived from comparing the time-and-space-averaged local "excitation velocities" (i.e., departure from the local circular Keplerian velocity) of all tracers of sizes $s_i$ and $s_j$. This fast and linear procedure cannot work in the much more complex context of debris discs, because of possible short timescale variations and the potential existence of different dynamical categories within a same size range. As a consequence, we perform a real-time estimate of all mutual $\Delta v_{i,j}=\left((Vim_x-Vjm_x)^2+(Vim_y-Vjm_y)^2+(Vim_z-Vjm_z)^2\right)^{0.5}$ values, where \vec{Vim} and \vec{Vjm} are \vec{Vi} and \vec{Vj} velocity vectors corrected from biases due to finite cell sizes. These biases are due to the Keplerian shear between tracers at different radial distances $r$ and the orientation of velocity vectors, because of different azimuthal positions $\theta$ within the cell, which has to be corrected by virtually rotating the tracers to the same longitude at the centre of the cell. In practice, to facilitate the correction process we switch to spherical coordinates. We first estimate, for each tracer, the departures $\vec{\delta V_i}$ and $\vec{\delta V_j}$ from the local Keplerian velocity. We then correct the magnitude of these two vectors by a factor accounting for the Keplerian shear, to obtain the debiased vectors $\vec{Vim}$ and $\vec{Vjm}$. This process corrects both for the difference in radial position within the cell and the azimuthal one.

\subsection{Collision outcome prescription}\label{colpresc}

The simplified 1\,m.s$^{-1}$ velocity barrier between erosion and accretion used for the protoplanetary disc case is clearly not adapted to the high-velocity destructive collisions regime of debris discs. We therefore implemented a collision outcome prescription as sophisticated as those commonly used in collisional evolution particle-in-a-box codes. We chose the energy scaling prescription described in \citet{theb07}, where erosive impacts are separated into two regimes, catastrophic or cratering, depending on the value of the specific impact energy per target mass unit $Q_{\textrm{\tiny{kin}}} = E_{\textrm{\tiny{col}}}/M_{\textrm{\tiny{t}}}$, where $E_{\textrm{\tiny{col}}} = 1/2 M_{\textrm{\tiny{p}}} M_{\textrm{\tiny{t}}} \Delta v^2 / (M_{\textrm{\tiny{p}}}+M_{\textrm{\tiny{t}}})$ (kinetic energy of the impact in the barycentric frame) as compared to the critical specific energy $Q^*$:
\begin{equation}
	Q^*=\alpha_1 \, \left(\frac{R_{\textrm{\tiny{tar}}}}{R_0}\right)^a+\alpha_2 \, \rho \,\left(\frac{R_{\textrm{\tiny{tar}}}}{R_0}\right)^b,
 	\label{Qstar}
\end{equation}
where $R_0=1\,$m, $M_{\textrm{\tiny{p}}}$ and $M_{\textrm{\tiny{t}}}$ are the respective mass of the projectile and target, and $\Delta v$ is the relative velocity between the two bodies colliding. The values for $a,b,\alpha,\beta$ can be found in the literature and depends on the physical composition of grains \citep[e.g.][]{hous90,benz99}. The value that have been used for this study are taken from \citet{benz99}: $a=-0.38$, $b=1.36$, $\alpha_1=3.5 \times 10^{3}$ J/kg, $\alpha_2=3 \times 10^{-8}$ (SI).

The first term in Eq.~\ref{Qstar} represents the strength regime, for which grains get harder to break as they get smaller ($a<0$). The second term, dominant for larger bodies, is the gravitational regime, for which bodies get harder to break as they get bigger ($b>0$). The code is able to cope with different chemical compositions of grains but for the sake of clarity we will assume pure silicates throughout the rest of this paper.

The fragmenting ($Q>Q*$) and cratering ($Q\leq Q*$) regimes are handled as described in the appendix of \citet{theb07}. For cratering impacts we do in particular distinguish between large-scale craters and grain-hitting-a-wall small impacts, connecting these two regimes by a smooth polynomial transition. The only simplification that we implement with respect to \citet{theb07} is that we only consider one power law (instead of two) for the size distribution of fragmented or cratered fragments. The index $p$ of the power law is constrained by the mass of the largest remaining fragment $M_{LF}$ coupled to the mass conservation condition and the constraint that there is only one body of mass larger than $M_{LF}$  \footnote{Note that we here implicitly assume that the largest fragment is part of the size distribution, so that the values for $M_{LF}$ and $p$ cannot be chosen independently and are interconnected. Some alternative prescriptions consider the largest fragment to be outside the power-law distribution. In this case, the coupling is between $M_{LF}$ and the \emph{second} largest fragment \citep[e.g][]{wyat02}}. This gives
\begin{equation}
	p = \frac{3}{-1 - \frac{N_{max} \, M_{LF}}{M_{tot}}},
 	\label{slopeq}
\end{equation}
where $N_\textrm{\tiny{max}}$ is the number of particles in the size bin of the largest fragment and $M_{tot}$ is the total mass of fragments.
$M_{LF}$ is given by the prescription of \citet{fuji77} for fragmenting impacts and by that of \citet{weth93} for cratering ones.

\subsection{Creation of collision-fragment "virtual tracers"} \label{virtual}

As already mentioned we are tracking all mutual impacts between tracers within each cell. For each tracer-tracer collision, we follow the procedure described in the previous section and estimate the total collisional mass lost by the impacting tracers. We then distribute this produced mass of collisional fragments into every size bin spanned by the $m<M_{LF}$ range, following the size distribution of index $p$. For every size bins that have received collisional matter, a new "virtual" tracer is created, carrying the amount of mass received. This tracer is produced at the position of the largest (that is, the one representing particles with the largest size) of the two impacting tracers and is given its velocity vector.

\subsection{Dynamical category sorting}\label{sorting}

At the end of the collision treatment routine, each spatial cell is thus populated with a vast number of "virtual tracers", carrying the mass of the fragments created by all tracer-tracer impacts, in addition to the remaining initial tracers that have lost some of their mass. The next and fundamental step of our procedure is to reduce this total number of tracers and regroup them into similar families for which only a few representative tracers are kept.

Identifying these families is a challenge in the context of the complex dynamics of debris discs, as LIDT-DD must be able to sort and regroup tracers not only as a function of physical size and spatial location (as in LIDT3D), but also according to their orbits, which can be very different for a given grain size at a given location (see Section \ref{specdd}). 
Note that for small, high-$\beta$ grains this variety of dynamical behaviours for a given grain size is an issue even in non-perturbed "quiet" discs. Indeed, at any given radial distance $r$, such grains can either have been produced locally, and thus have their periastron $q$ close to $r$, or much further closer to the star, with $q<<r$ and thus a very different orbit, and reach $r$ because of the highly-eccentric orbit imposed by radiation-pressure.

The sorting method has to be generic enough so that no special treatment will be required when applied to different cases (unperturbed discs, planet/disc interactions, transient massive break-ups, $\ldots$). We chose the hierarchical cluster analysis of \citet{ward63}, which satisfies this requirement and has the advantage of not requiring the user to define in advance the number of different dynamical populations. This procedure looks for distinct groups, as defined by the distribution of mutual "distances" between tracers in a multi-parameter phase space. The two parameters that we consider for the cluster analysis are the tracers' periastron $q$ and apoastron $Q$, which are enough to constrain a particle orbit present in a ($r,\theta$) spatial cell. In practice, the dynamical family identification is performed in a $q+a$ vs. $Q-a$ plane, instead of simply $q$ vs. $Q$ one, in order to give the same weight to $q$ and $Q$ when looking for "distances" in the parameter phase space. Note that the longitude of the periastron $\omega$ is not needed for the sorting, because the limited azimuthal extension of the spatial cell implicitly confines the values for $\omega$ within a narrow range once the $q$ and $Q$ sorting has been done. Likewise, the particles' angular position on their orbit is almost fully implicitly constrained by the narrow radial extension of the cell, the only two possible distinct configurations being particles on their way "out" (i.e., moving towards apoastron) or "in" (moving towards periastron). To distinguish between these two categories, we simply sort tracers into two subcategories depending on the sign of their radial velocity $v_r$.
The only free parameter for the sorting is the precision criteria, which determines how different two populations need to be to be considered as dynamically distinct. To adjust this criteria we chose the trial and error approach by performing test runs with an unperturbed disc and chose the largest possible value that could reliably sort high-$\beta$ particles as a function of their different production annulii. Note that this procedure bears some similarities with the "closest-streamline-search" of \citet{star09}, except for the crucial difference that our procedure is able to create new, non pre-existing dynamical families as the system evolves.

\begin{figure}
   \centering
     \includegraphics[width=9cm]{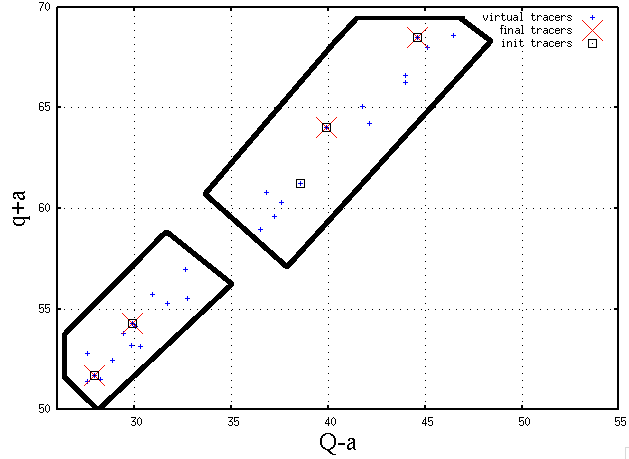}
      \caption{Dynamical family sorting procedure. ($q+a$) vs. ($Q-a$) for all tracers, corresponding to the smallest size bin ($\beta=0.44)$, which are present in one given spatial cell. We plot the tracers present at the beginning ("init tracers") of a collisional time step, plus the "virtual tracers" that are temporarily created as a result of mutual tracer-tracer collisions amongst larger size bins, and the "final tracers" that are eventually kept at the end of the sorting and selection procedure. 
The thick black lines show the two dynamical categories identified by the sorting algorithm (see Section \ref{sorting}). The set-up is that of the pedagogical test run, of an unperturbed disc initially confined to an inner ring of parent bodies, presented in Section \ref{illu}. The considered spatial cell is located 8\,AU outside the inner birth ring.}
         \label{figsorting}
   \end{figure}

The procedure is summed up in Fig.~\ref{figsorting}. This plot shows, for a given spatial cell and a given size bin (corresponding to small grains with $\beta=0.44$), the tracers' orbital distribution, in a $q+a$ vs $Q-a$ plane, before ("init" tracers) and after ("final" tracers) a collisional time step. We also plot the "virtual" tracers that are temporarily created as a result of collisions amongst tracers corresponding to larger size bins. As can be seen, some of these virtual tracers will be kept as "final" tracers by the sorting procedure, while some will be discarded and their mass transferred to the selected final tracers from  their dynamical family (see Sec. \ref{reassignment}). In order not to loose any dynamical information, tracers that were present at the beginning of the time step ("init" tracers) have priority over "virtual" tracers to be selected as "final" tracers. The set-up we considered is the one corresponding to the pedagogical run presented in Sec. \ref{illu}, with an unperturbed disc initially confined to a narrow ring around 12 UA. The spatial cell that is considered in Fig.~\ref{figsorting} is located $\sim 8\,$AU beyond the ring and is thus mostly populated with small grains whose orbits stretch to these regions because of their high-$e$ imposed by radiation pressure. These grains originating from the inner ring constitute one of the dynamical families (delimited by the thick black lines) identified by the sorting procedure, i.e., the one in the lower-left region of the plot (smaller $q$ and $Q$). The second dynamical family, with larger periastron, corresponds to grains that have been produced locally, as fragments from collisions involving slightly larger particles.

\subsection{Tracer reassignment}\label{reassignment}

As soon as all dynamical categories have been identified amongst all initial and virtual tracers present in a cell, we then eliminate all "redundant" tracers within these families, i.e., only keeping a small number of representative tracers carrying all the mass of their category.

Given the potentially large number of dynamical categories for each particle size at each given location, and to avoid an unmanageable number of tracers, we only keep a maximum number of 2 tracers per dynamical family per size bin per spatial cell. To select these two representative tracers, we give priority to the "initial" tracers, i.e., those that were already present at the beginning of the time step before the collisional stage, and either select 2 tracers at random amongst them or, if there is only one such tracer, select this tracer and choose the other at random amongst the "virtual" tracer pool. In the case when there is no initial tracer at all within a given family, then the two representative tracers are directly chosen at random amongst the virtual population.

\begin{figure}
   \centering
     \includegraphics[width=9cm]{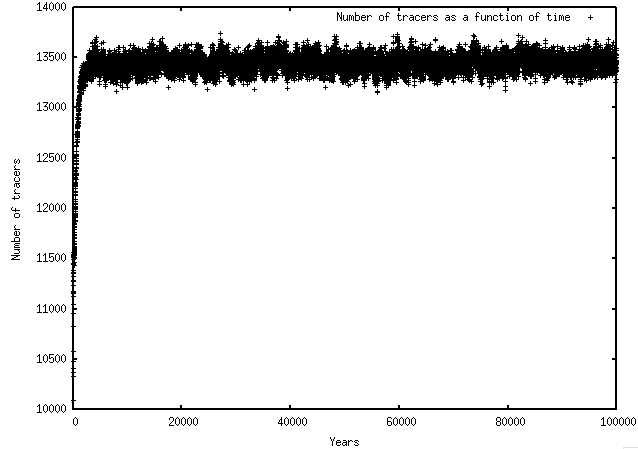}
      \caption{Fiducial inner-parent-body-ring run (see Section \ref{illu}). Evolution of the total number of tracers as a function of time. }
         \label{fignbtrac}
   \end{figure}

Once the two representative tracers have been selected, all other tracers within the same family are discarded and their mass is evenly added to each of the two final tracers.
All these discarded tracers are turned off and stored to be used at another time step in other cells when they will be needed. As illustrated by Fig.\,\ref{fignbtrac}, this sorting and selection procedure effectively limits the increase of the number of active tracers after an unavoidable initial adjustment period.

\subsection{Collisional energy dissipation}\label{colldiss}

High-velocity collisions necessarily alter the orbits of impacting bodies and their resulting fragments by redistributing and dissipating kinetic energy. In simplified collision models where bodies are treated as inelastically bouncing hard spheres, the kinetic energy dissipation is usually modelled by a normal and a tangential restitution coefficients, $\epsilon_{N}$ and $\epsilon_{T}$, for the relative velocity, whose standard values are $\epsilon_{N} = -0.3$ and $\epsilon_{T}=1$ \citep[e.g.,][]{theb98,marz00,char01,lith07}. Such a simple prescription is impossible to directly implement here given the complexity of the tracer creation, sorting and reattribution procedures. However, it is possible to post-process tracer orbits to account for the average energy dissipation induced by collisions. 

In practice, at the end of our collision procedure, once all final tracers have been selected within a spatial cell, we run an additional procedure that treats these tracers as potentially bouncing hard spheres. For each pair of tracers $i$ and $j$, we first compute mutual collision probabilities the same way as in Sec.~\ref{collt}. Using these probabilities, we then randomly select tracer pairs that will effectively get their velocity vectors modified by inelastic collisions during timestep $dt$. For these chosen pairs of tracers, we then assume that an inelastic collision occurs between two particles having the physical sizes that these tracers stand for. This collision is treated in the centre of mass frame and each body's velocity modified following the $\epsilon_{N}$ and $\epsilon_{T}$ prescription the same way as in \citet{theb98} and \citet{char01}. This procedure is of course not an exact estimation of the post-collision evolution of each individual tracer present in a cell, since a given tracer will have its velocity vector modified by this procedure typically every $\sim t_{coll}/dt$ timesteps. However, on timescales exceeding a few $t_{coll}$ it does accurately model the average energy dissipation induced by collisional activity, and reproduces all the expected behaviours predicted for simple cases (see Sec.~\ref{dissip})

\subsection{Illustrative test run}\label{illu}

\begin{figure*}
\makebox[\textwidth]{
\includegraphics[scale=0.25]{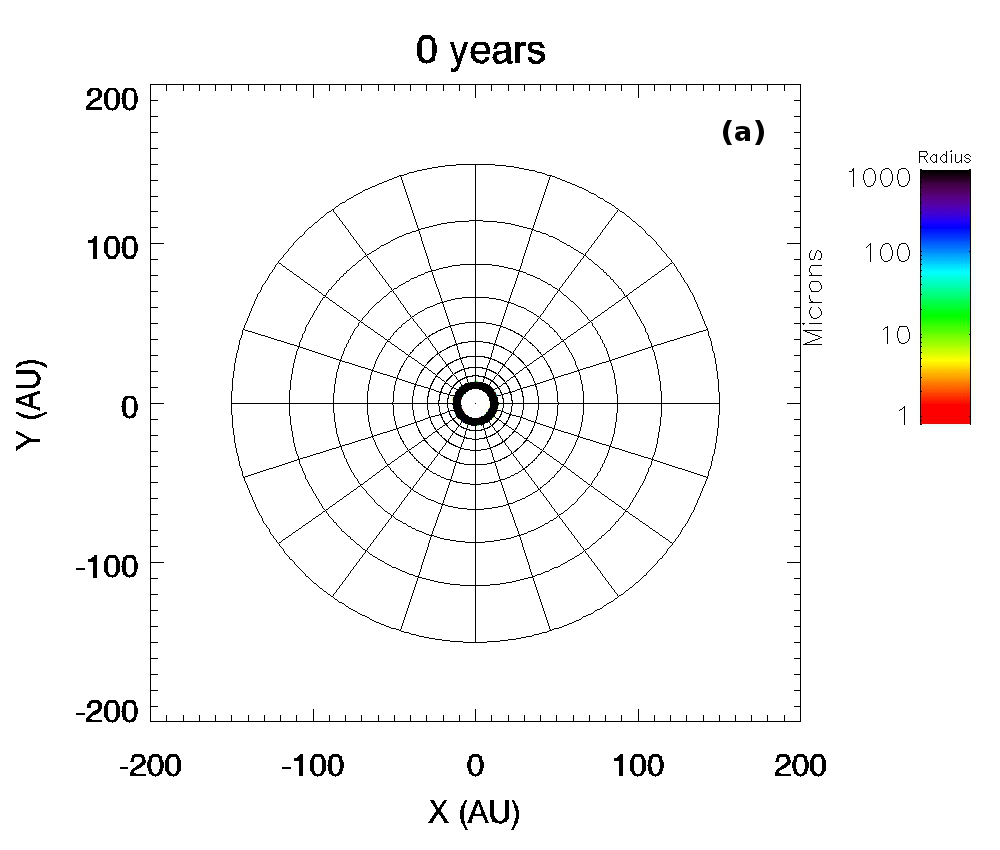}
\includegraphics[scale=0.25]{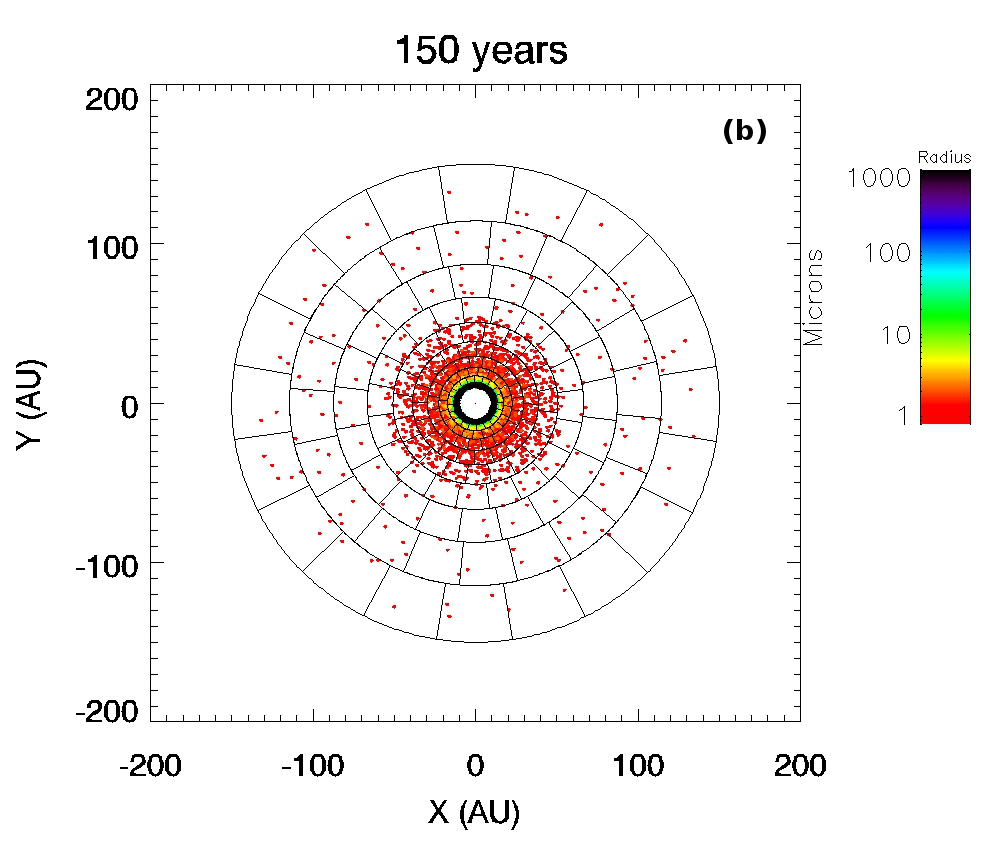}
}
\makebox[\textwidth]{
\includegraphics[scale=0.25]{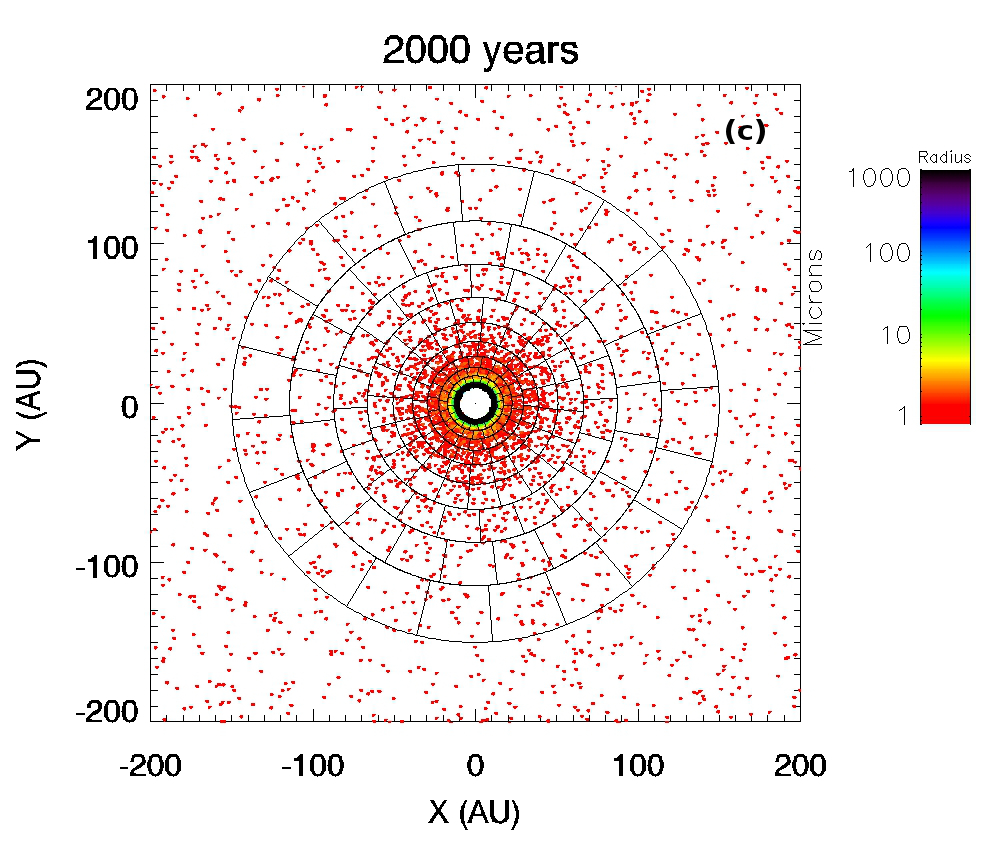}
\includegraphics[scale=0.25]{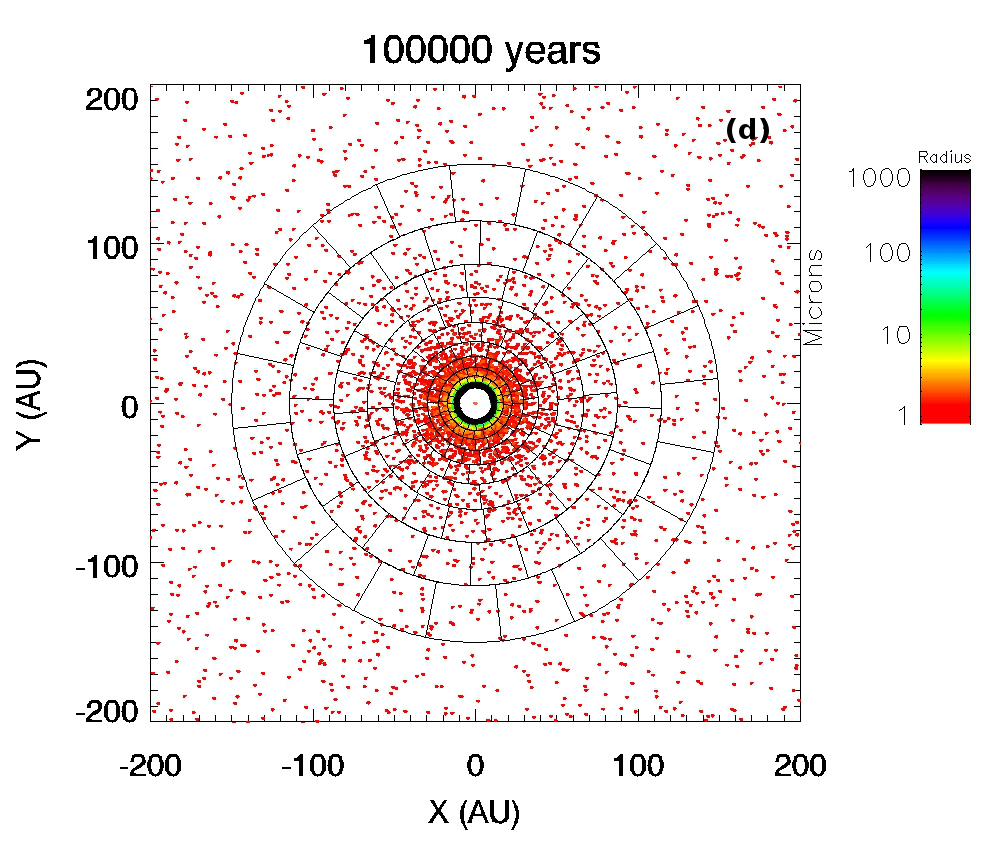}
}
\caption[]{Fiducial inner-parent-body-ring run (see Section \ref{illu}). 2-D distribution of all tracers at different epochs. The colour scale indicates the size bins (in $\mu$m) represented by each tracer (see text for details).}
\label{simutest}
\end{figure*}

To better illustrate the way the code functions, we present a simple pedagogical run, considering the collisional evolution of an unperturbed debris disc. The system's evolution is presented in Fig.~\ref{simutest}, showing snapshots of the tracer's distribution at different times.

We start from a narrow ring, between $a_\mathrm{min}=11$ AU and $a_\mathrm{max}=12$\,AU, where all the matter is initially confined (Fig.~\ref{simutest}a). The ($r,\theta$) "collisional grid", where the collisional evolution of solids is treated (see Section \ref{colpresc}) extends out to 150\,AU.
This grid is made of 10 radial and 20 azimuthal cells from 10 to 150 AU and a log-scale is used for the radial spacing between the cells. The grid is rotating differentially so that each ring (at same $r$) rotates at its local keplerian velocity. 
The outer edge beyond which the dynamical evolution is not followed is set at $r_\mathrm{outDyn}$=300\,AU. We consider a $\beta$ Pic-like A5V star and pure compact silicates for all solids in the disc. The size distribution of this population of solids is followed from 50\,km down to the radiation-pressure cut-off size at $2\,\mu$m. This size range is divided into size bins for which the logarithmic spacing is 1.15 for the critical domain of small grains in the $2\,\mu$m to 0.1mm range, and 2 for larger solids (see Section \ref{sampling}).
The total initial mass of the disc is $7\times10^{24}$\,kg, corresponding to an average optical depth $<\tau>\, \sim 10^{-3}$ in the birth ring, and is distributed following a differential power law in $s^{-3.5}$ from $s_{max}$ to $s_{min}$.
All the main parameters for the considered setup are summarized in Tab.~\ref{tabtest}.

\begin{figure}
   \centering
     \includegraphics[width=9cm]{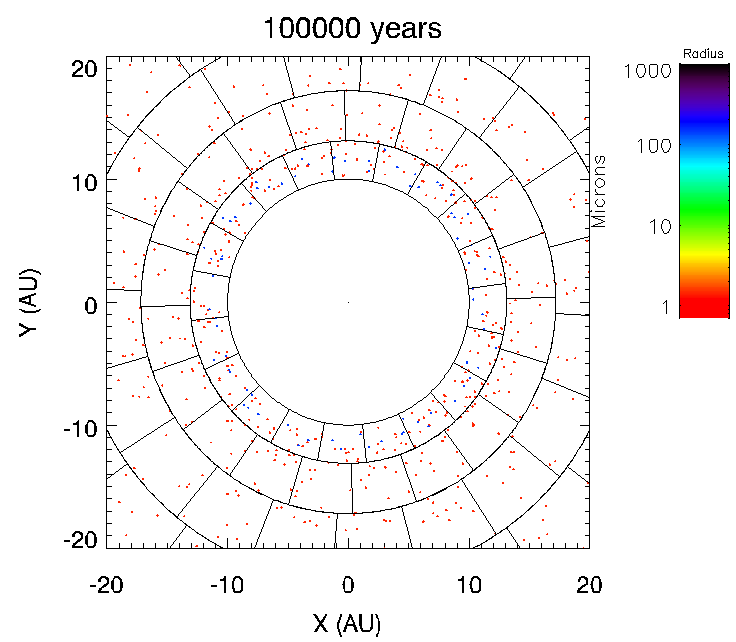}
      \caption{Fiducial inner-parent-body-ring run (see Section \ref{illu}). Tracer distribution at $10^{5}\,$years. Only tracers for two size bins are shown, $\beta=0.4$ (in red) and $\beta=5\,10^{-3}$ (in blue).}
         \label{figtwobins}
   \end{figure}

In the earliest stage of the disc evolution, we see the tracers corresponding to the smallest grains moving out from the birth ring because of radiation pressure that places them on high-$e$ orbits (Fig.~\ref{simutest}b). This outward movement of the initial tracers leaving the ring is compensated by the creation of new tracers in the ring due to impacts amongst larger particles. As a consequence, the total number of tracers is initially increasing (see Fig.~\ref{fignbtrac}). The increase stops when the small-grain tracers produced in the ring have had enough time to come back in the ring after having travelled through their elongated orbits. The number of tracers thus reaches a plateau after a few dynamical timescales of the smallest grains, i.e., a few $10^{3}$ years. After this point, no significant change is visible when looking at the spatial distribution of the \emph{tracers}, since each "collisional cell" is populated with approximately 2 tracers per dynamical category per size bin (see Section \ref{sorting}). The importance of these dynamical categories and of the dynamical class sorting procedure is clearly illustrated in Fig.\,\ref{figtwobins}, showing the spatial distribution of tracers corresponding to a very small size bin ($\beta$=0.4) as well as to the thirtieth size bin ($\beta = 5\times10^{-3}$), the latter corresponding to larger particles not affected by radiation pressure. As can be easily seen, the "big" tracers are logically confined to the innermost cells of the grid while the "small" tracers populate the whole system. In addition to that, while the number of big tracers is on average close to 4 per collisional cell (2 for those moving towards apoastron and 2 for those on their way back), the number of small grain tracers is much larger. This is due to the fact that, for these small grains, there are, for the same given spatial location, several dynamical behaviours that are possible depending on where the grains have been produced (see Section \ref{specdd}).

\begin{figure*}
\makebox[\textwidth]{
\includegraphics[scale=0.25]{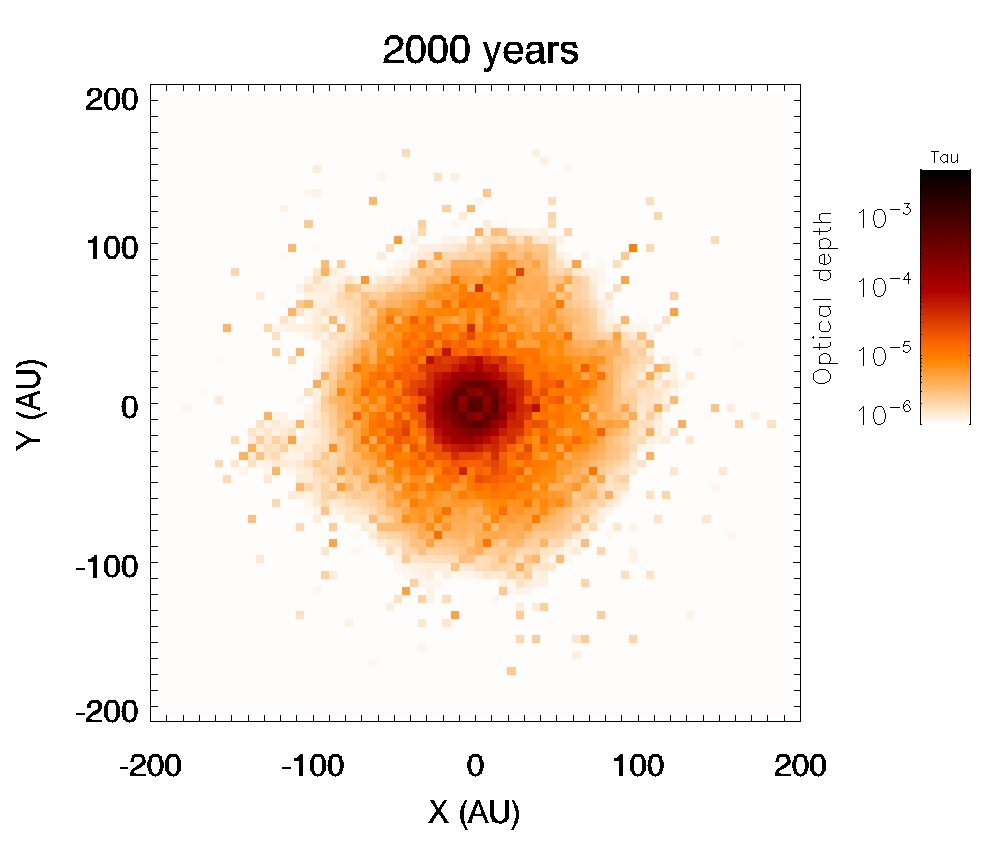}
\includegraphics[scale=0.25]{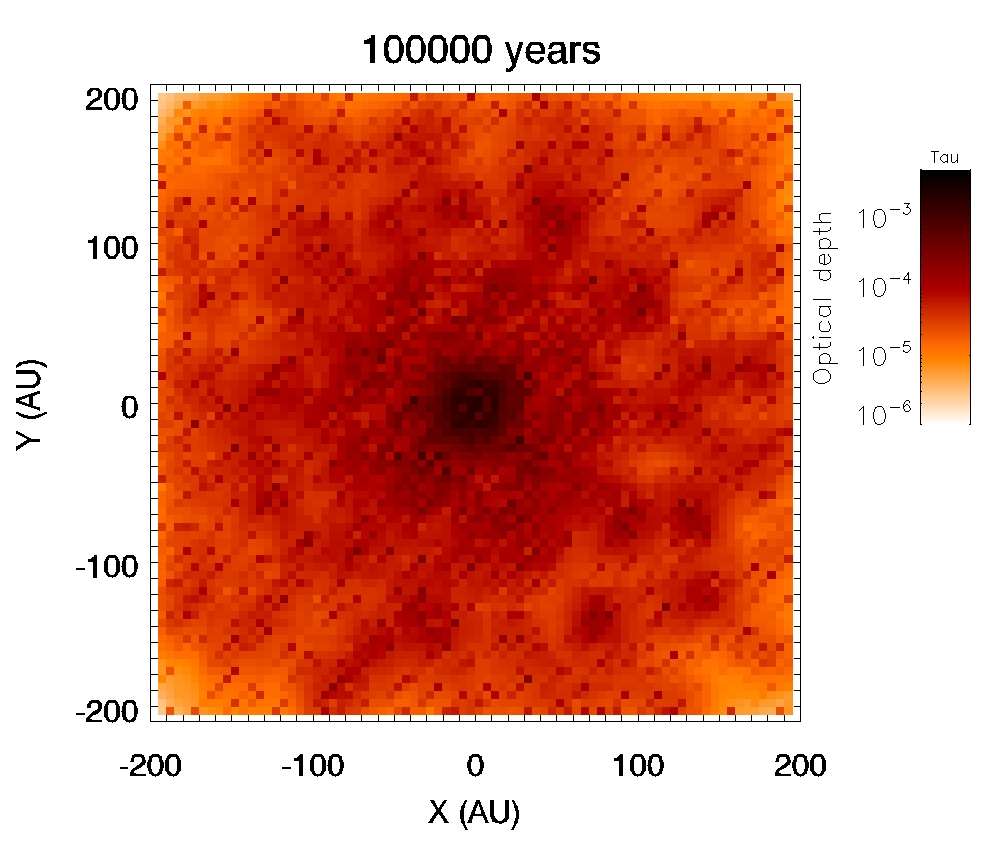}
}
\caption[]{Fiducial inner-parent-body-ring run. 2-D geometrical optical depth map after 2000\,years (left) and $10^{5}\,$years (right). Note how the maps greatly evolve between these two epochs, while the distribution of tracers globally stays the same (Fig.~\ref{simutest}c \& d).}
\label{simutestoptdepth}
\end{figure*}

It is important to stress that the system has not yet reached a steady state by the time the number of tracers has reached its plateau (a few $10^{3}\,$years). It will continue to evolve because of collisions, which progressively change and redistribute the amount of mass carried by each tracer. So that even if the global distribution of \emph{tracers} no longer changes after $\sim 2\times10^{3}$\,years (Fig.~\ref{simutest}c \& d), the corresponding 2-D maps of the disc's \emph{optical depth} is still evolving long past this point (Fig.~\ref{simutestoptdepth}a \& b).
The steady state, for which the optical depth maps no longer evolves, is reached only after $\sim 10^{5}$\,years.

The computation time to reach $10^{5}\,$years evolution using the OPEN-MP version of the code on 8 CPUs is $\sim 65$ hours. The collisional procedure is by far the most costly in terms of CPU time. More precisely, for one typical time step the dynamical evolution procedure (to work out new orbits) takes up only $\sim 1$\% of the total computation time, whereas the collisional procedure uses the remaining 99\%. Within this collisional procedure, $\sim 1.5$\% of the time is used to compute all tracer-tracer encounters and impact velocities, $\sim 73$\% to compute the outcome (fragments) of these collisions, $\sim 24$\% to sort out the dynamical families for all tracers and $\sim 1.5$\% to select all the "final" tracers that will be kept at the end of the time step and to reassign the mass onto these final tracers.

Note that we chose this simple unperturbed example because it has the pedagogical virtue of allowing to easily distinguish the dynamical and collisional evolution of the tracers, i.e., the spatial distribution of tracers stabilizes long before their collisional evolution does. In more complex cases (the ones that are really interesting to investigate with this code) such easy distinction cannot be made, because the spatial distribution \emph{and} the collisional evolution of tracers always change simultaneously (see for example the case study considered in Section \ref{hvc}).

\begin{table}[h]
\caption{\label{tabtest}Relevant parameters used for the fiducial test run simulation}
\centering
\begin{tabular}{lc}
\hline\hline
Star\\
\hline
Solar type & A5V \\
Mass        & $1.7 M_\odot$\tablefootmark{a}\\
Radius    &  $5 R_\odot$ \\
\hline
Grains\\
\hline
Blowout size ($s_\mathrm{cut}$) & $2.06 \, \mu m$ \\
Material & Silicate \\
Porosity & 0 \\
Density ($\rho$) & 3000 $\textrm{kg.m}^{-3}$ \\
\hline
Debris ring population\\
\hline
Material & Basalt \\
Minimum size ($s_\mathrm{min}$) & $2.06 \, \mu$m ($\beta$ = 0.44) \\
Maximum size  ($s_\mathrm{max}$) & 58 km  \\
Initial radial extent & $11 < a < 12 $ AU\\
Initial eccentricity & $0 < e < 0.1$ \\
Initial size distribution & $r^{-3.5}$ Dohnanyi power law \\
Size sampling & $\epsilon_f=1.15$, $\epsilon_c=2$ \\
Total mass ($M_\mathrm{tot}$) & $7 \times 10^{24}$ kg \\
Number of initial tracers & 20 000 \\
Optical depth & $<\tau> \,= 1 \times 10^{-3}$ \\
\hline
Collisional prescription\\
\hline
Q* prescription & see Sec.~\ref{colpresc} \citep{benz99} \\
Threshold energy (1 cm)      & $Q_0^* = 10^3$ J/kg \\
Power law index & p (see Eq.~\ref{slopeq}) \\
Cratering & 3 laws (see section \ref{colpresc}) \\
Fragmentation & \citet{fuji77} \\
\hline
\end{tabular}
\tablefoot{Simulation parameters used for the illustrative fiducial inner-parent-body-ring test run.\\
}
\end{table}

\section{Tests}

Testing LIDT-DD is a challenge, as this code is the first of its kind, and there do not exist reliable results regarding the coupled dynamical and collisional evolution of debris discs that can be used as references. There exist, however, simplified cases for which robust results have been obtained in past studies, which can be used as a benchmark to test the different aspects of our code.

\subsection{Conservation of angular momentum}

\begin{figure}
   \centering
     \includegraphics[width=9cm]{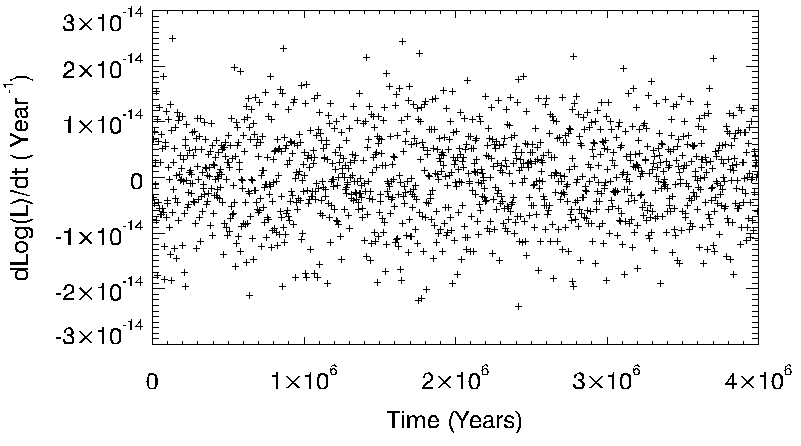}
      \caption{Inner-parent-body-ring test run (see Section \ref{illu}). Angular momentum derivative variations over time (dLog(L)/ dt).}
         \label{figangmom}
   \end{figure}

In the absence of external perturbers, the disc's angular momentum needs to be conserved in a problem where all the forces are central (we do not consider PR-drag in this section). Fig.~\ref{figangmom} presents the evolution of the angular momentum derivative (dLog(L)/dt) for a few million years of our illustrative test run of an unpertubed ring (see Sec.~\ref{illu}). As can be seen, there are unavoidable small stochastic variations on short timescales, due to the tracer selection and reattribution procedure. These variations remain, however, very limited, less than $2\times10^{-14} \,\textrm{Yr}^{-1}$ in relative amplitude. More importantly, these stochastic variations do not increase in amplitude and induce no drift of the angular momentum over long timescales.

\subsection{Mass loss}

The collisional grinding of a debris disc naturally removes mass from it because of the blow-out of the smallest grains by radiation pressure. For the case of an unperturbed system left to itself, the expected temporal evolution of both the system's total and dust masses has been investigated in numerous studies. For an idealized system where the collisional cascade has had enough time ($t > t_{max}$) to reach the largest bodies in the size distribution, the expected behaviour is a decrease of $M_{tot}$ and $M_{dust}$ that is $\propto t^{-1}$ \citep{domi03,wyat07}. However, this asymptotic behaviour is only expected to be reached at very late times, which can be longer than a system's age \citep[e.g.,][]{lohn08}. For a more realistic case where $t < t_{max}$, detailed numerical and analytical investigations have shown that the evolution of $M_{tot}$ and $M_{dust}$ is much more complex, in particular because of the size dependency of the $Q*$ parameter in both the strength and gravity regimes \citep{lohn08,wyat11,gasp13}.
We compute both $M_{tot}$ and $M_{dust}$ for our test case of an unperturbed disc and compare our results to that of these earlier studies.

\begin{figure}
   \centering
     \includegraphics[width=9cm]{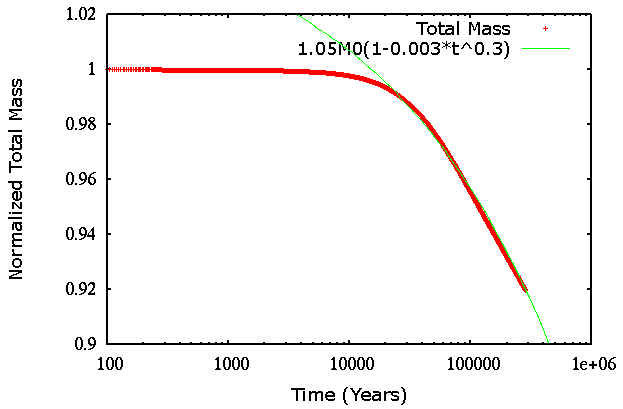}
      \caption{Inner-parent-body-ring run (see Section \ref{illu}). Evolution of the normalized total mass of the system.}
         \label{figmassevol}
   \end{figure}

The normalized total mass evolution displayed in Fig.~\ref{figmassevol} does closely match the behaviour obtained and thoroughly analysed by \citet{lohn08}: During an initial stage, for which the collisional cascade has reached a steady-state only for bodies in the strength regime, $M_{tot}$ stays almost constant. This phase ends when $t\geq t_{b}$, where $t_{b}$ is the time at which the collisional cascade has reached objects that are large enough for $Q*$ to be in the gravity regime. After that, $M_{tot}$  decreases faster, at a rate which closely matches that predicted by Equ.~39 of \citet{lohn08}. If we indeed substitute in this equation our own parameters for the initial size distribution and the $Q*$ dependency, then we get $M_{tot} \sim M_0(1-0.0030\, t^{0.3})$. As can be seen, this behaviour is a very good fit to our test run at later times. Note that our best fit is obtained for $1.05M_0$ instead of $M_0$, but this small difference is expected, as L\"{o}hne's formula neglects all mass evolution during the $t<t_b$ period.

\begin{figure}
   \centering
     \includegraphics[width=9cm]{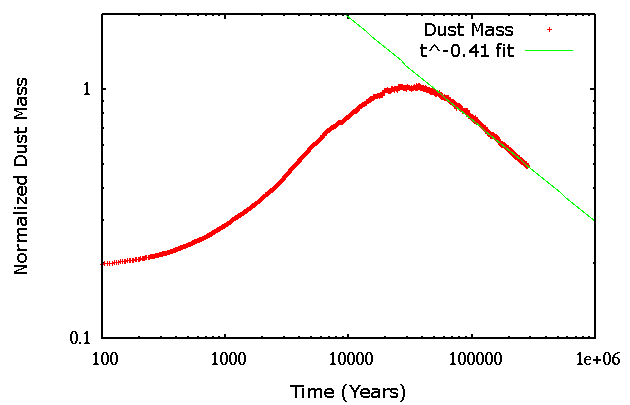}
      \caption{Same as Fig.~\ref{figmassevol} but for total $dust$ mass (all bodies $\leq 1\,$mm).} 
         \label{figmassevoldust}
   \end{figure}

As for the dust mass (Fig.~\ref{figmassevoldust}), it first increases during the initial $t\leq t_b$ phase. This is an expected result, due to the fact that the equilibrium size distribution of particles in the strength regime is steeper than the initial PSD \citep{gasp13}. Beyond $t_b$, $M_{dust}$ falls off approximately as $\propto t^{-0.41}$, which is again fully compatible with the rate predicted by \citep{lohn08}, i.e., between $t^{-0.3}$ and $t^{-0.5}$ (see Fig.10 of that paper).

\subsection{Collisional energy dissipation} \label{dissip}

  \begin{figure}
   \centering
     \includegraphics[width=9cm]{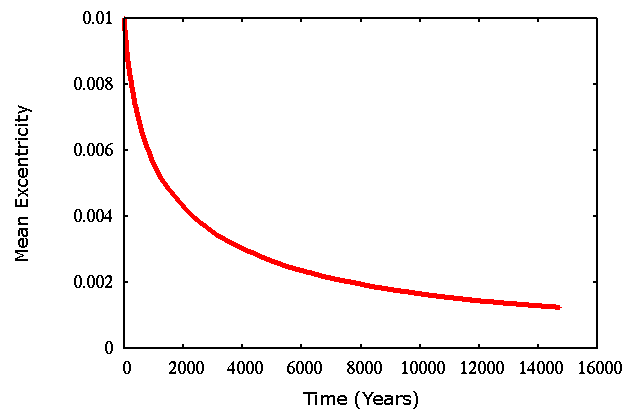}
      \caption{Mean eccentricity evolution over time for an unperturbed narrow ring centered at 1AU with a narrow size distribution and $\tau_0=0.02$ (see text for more details) }
         \label{figexmoy}
   \end{figure}

There is to our knowledge no reference analytic expression for how the kinetic energy should dissipate for complex systems where collisions and dynamics are coupled. Even for an unperturbed disc, there is no available law for high-velocity fragment-producing collisions affecting an extended size distribution, especially when taking into account radiation pressure.
We can, however, test our code against available results obtained by \citet{lith07} for the simplified case of a ring made of monosize particles large enough not to be affected by radiation pressure. To mimic such a monosize-particle case with LIDT-DD while retaining our collision-outcome prescription and tracer creation routine, we consider a system with 10 size bins, but confined within a relatively narrow spread in sizes (all material that goes below the smallest bin is labelled as "lost dust").

We take the same set-up as the nominal case considered by \citet{lith07}, i.e., a narrow ring at 1 AU from the central star, a mean tracer eccentricity of 0.01 and an optical depth $\tau_0 \sim 10^{-2}$. We also chose the standard values $\epsilon_N=-0.3$ and $\epsilon_T=1$ for the collision restitution factors. As an indicator of how the energy evolves in the system we chose to follow the evolution of the average eccentricity of tracers in the disc.

The $\left< e(t) \right> $ evolution we obtain (Fig.~\ref{figexmoy}) is very close to the one obtained by \citet{lith07} in their Fig.~1. for their "hot" case, with for example an initial decay time of $\sim 2.5\times 10^{3}$ years to reach $e_0/2$.

Another important result from this graph is that the tracers' dynamical evolution does not suffer from an artificial heating due to the finite cell sizes. Indeed, the values of $<e>$ that are reached are far below the ratio $\Delta\,r_{cell}/r\sim 0.25$ between the width of the spatial cells and their radial distance. This proves the validity of our Keplerian-shear and azimuthal correction procedures when estimating relative velocities and collision outcomes between tracers of the same cell.

\subsection{Viscous spread} \label{viscous}

Another consequence of the kinetic energy redistribution after collisions is the viscous spread of the disc, dissipative collisions playing here the role of viscosity \citep[e.g.,][]{lynd74}. For an unperturbed narrow ring, analytical derivations predict that the radial width of the ring should expand as $\propto t^{1/3}$ \citep{lith07}. We check this behaviour for the no-radiation-pressure and narrow-size-distribution case considered in the previous subsection. As the typical timescale for the ring's expansion scales as $\Delta^{3}$ (where $\Delta$ is the ring width) and $\tau_{0}^{-1}$, and can thus be very long for wide and/or tenuous discs, we follow \citet{lith07} and consider a very narrow and dense initial ring, centered at 10\,AU, of initial width $\Delta_0=0.05$ AU. We consider the same "optimum" configuration as in \citet{lith07}, i.e., an initial eccentricity distribution such as the mean radial excursion of the tracers is equal to $\Delta_{0}$.

\begin{figure}
   \centering
     \includegraphics[width=9cm]{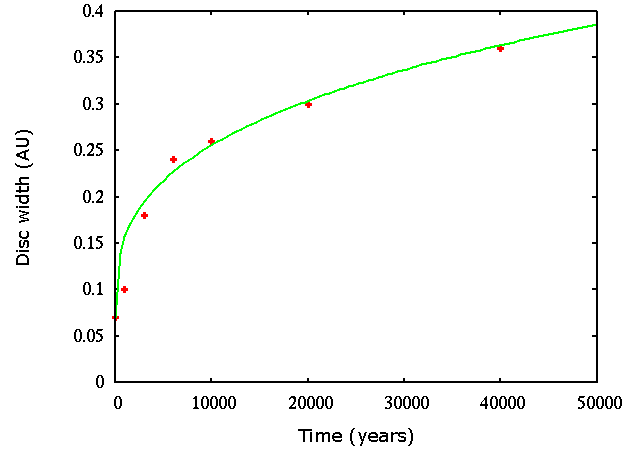}
      \caption{Viscous spreading: evolution of the radial width of an unperturbed, initially very narrow and very dense ring \citep[as in][]{lith07}. The red crosses indicate the ring's full width at half-maximum measured at different epochs. The green curve gives the theoretical behaviour in $\Delta = \Delta_0(1+C.t^{1/3})$, for the best-fit value $C=0.13$ }
         \label{figviscsp}
   \end{figure}

As can be seen in Fig.~\ref{figviscsp}, for a fixed initial set-up, the evolution of the ring's width does closely match the expected behaviour in $t^{1/3}$. This is exactly the behaviour expected for a narrow ring evolving under dissipative collisions. Moreover, we can also check that the magnitude of the spreading is fully compatible with viscosity being physical (i.e., collisional) rather than numerical. To do so we follow \citet{lith07} and consider the simplifying assumption that the typical time for a disc to diffuse the width $\Delta$ is

\begin{equation}
t_{spread} \sim t_{col}\,\,\left(\frac{\Delta}{<e>.\,r}\right)^{2},
\label{eqdiff}
\end{equation}

where $< e >$ is the average eccentricity of particles in the ring, and $r$ the ring's radial distance to the star. This equation is valid as soon as $\Delta$ is greater than $<e>.\,r$. Replacing $t_{col}$ by its approximate value $t_{orb}/(2\pi \tau)$, where $\tau$ is the optical depth in the ring when it has reached width $\Delta$, we get
\begin{equation}
t_{spread} \sim \frac{t_\textrm{\tiny{orb}} \Delta^3}{2\pi \tau_0 \Delta_0 (<e>.\,r)^2}
\label{eqlith}
\end{equation}
where we used $\tau \sim \tau_0 (\Delta_0/\Delta)$. 
Plugging in the parameters considered for the present set-up ($r=10\,$UA, $\Delta_0=0.05\,$UA, $\tau_0=0.01$), we compare estimates given by Equ.~\ref{eqlith} with our numerical results displayed in Fig.~\ref{figviscsp}. We find for example that the typical time to reach $\Delta = 0.25\,$UA should be $\sim 12000\,$years, which is relatively close to the $\sim 10000\,$ retrieved from Fig.\ref{figviscsp}. Likewise, the time to reach $\Delta = 0.37\,$ UA should be $\sim 34000\,$years, once again relatively close to the $\sim 40000\,$ obtained in our run.

Note that for this ultra-narrow/very-dense ring case, the spreading effect is much stronger than it is for the "nominal" fiducial ring considered elsewhere in this study (see Sec.~\ref{illu}). This is fully logical and is due to the fact that our nominal ring is both much wider, $\sim 1\,$AU instead of 0.05\,UA, and more tenuous, with $\tau_0=10^{-3}$ instead of $0.01$\footnote{Note that the "effective" $\tau'_0$ of the particles for which viscous spreading should be the easiest to identify in our nominal run is even much less than these $10^{-3}$. Indeed, $\tau_0$ is dominated by very small grains which are placed by radiation pressure on very eccentric orbits, whose radial excursion largely exceeds (and masks) the effect of viscous spreading. The optical depth contained in particles large enough to be only weakly affected by radiation pressure is lower than $\tau_0$. And the optical depth contained in the larger bodies that control the collisional cascade, and thus the location of the mass reservoir for the ring's evolution, is orders of magnitude lower than $10^{-3}$.}.

\subsection{Outward and inward flow of mass}

One of the most innovative aspects of the code is the identification and sorting of dynamical categories within tracer populations and the subsequent tracer-reassignment procedure. As described in Sections \ref{sorting} and \ref{reassignment}, these two procedures are relatively complex and they together constitute one of the most critical part of LIDT-DD. 

A test for the reliability of these procedures can be made with the simplified set-up considered in Section \ref{illu} of an unperturbed inner ring of parent bodies. In this case, in regions outside the main ring there should be a 2-way flow of small, radiation-pressure affected grains produced in the annulus and placed on high-e orbits by radiation pressure: one flow of particles moving outward towards their apoastron and another flow of particles moving inwards on their way back to their periaston in the ring. At steady state, for a given particle size $s_0$, there should be an almost perfect balance between these two flows, with only a small excess of outbound grains due to the small amount of $s_0$ grains that have been collisionally destroyed during their stay in the outer regions on their elongated orbits. In a fully deterministic (and fully unachievable) code where all numerical particles correspond to real physical bodies, this behaviour should be obtained automatically. Here, however, this behaviour is far from being straightforward, because particles are regrouped into tracers that can in principle be reshuffled, created, destroyed and/or merged at every collisional time step, depending on the number of dynamical categories that are identified and the number of redundant tracers within each of them.

\begin{figure}
   \centering
     \includegraphics[width=9cm]{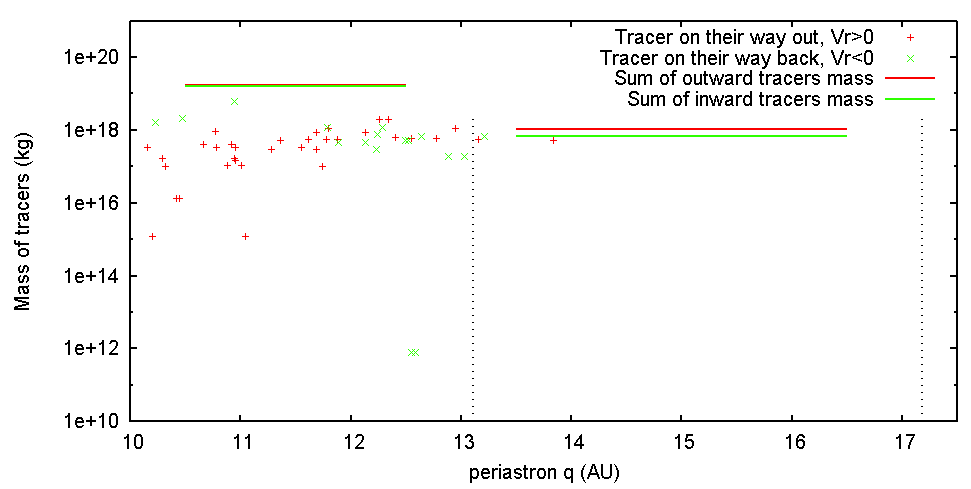}
      \caption{Inner-parent-body-ring run at quasi steady-state (see Section \ref{illu}). Mass carried by tracers corresponding to the $\beta=0.4$ size bin, located in the fifth radial ring of the spatial grid (between 29 and 39 AU). Tracers are plotted as a function of their periastron. Tracers in red are moving outward whereas those in green are moving inward. The black dotted vertical lines delimit the radial extension of the first and second radial annulii of the spatial collisional grid. The horizontal red and green lines correspond to the total mass carried by all outbound and inbound tracers, respectively, having their periastron in these first and second annulii.
}
         \label{figmq}
   \end{figure}

We check if this result is obtained by displaying, in Fig.~\ref{figmq}, the mass contained in all tracers corresponding to the $\beta=0.4$ size bin, which are present, at a given time, in all the azimuthal cells located in the fifth radial ring between 29 and 39 AU. We plot these tracers as a function of the periastron of their orbit, which gives the region of origin of the grains they represent.
For tracers having their periastron in one given radial region, we see that, although there are significant variations in the mass carried by individual tracers, the \emph{total} masses carried by outbound and inward tracers (green and red lines) are remarkably close, with only the expected small excess in favour of outbound grains. The destruction rate, numerically estimated to be of the order of $10^{10}$ kg per time step is so low (in outer parts of the disc $\tau < 10^{-4}$) in this simulation that the red and green lines should be, and are almost superposed.

\subsection{Shape of the Size Distribution}

Over the past decade, statistical particle-in-a-box models have identified several robust results regarding the shape and profile of the Particle Size Distribution (PSD) of collisional debris discs, in particular how it might depart from the standard equilibrium distribution in $s^{-3.5}dr$ \citep{dohn69}. We verify that all these robust features are obtained when our code is applied to unperturbed discs similar to those considered in those past investigations.

\begin{figure}
   \centering
     \includegraphics[width=9cm]{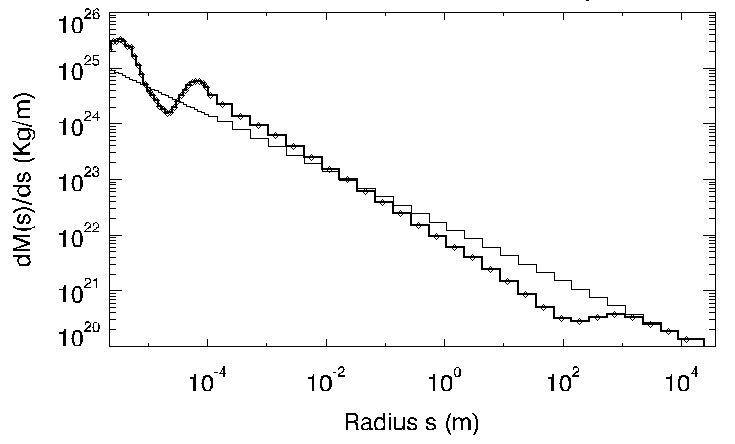}
      \caption{Inner-parent-body-ring run (see Section \ref{illu}). Differential particle size distribution, at steady state, expressed in terms of dm/ds, in the main inner ring. The thin black line corresponds to a distribution in $s^{-3.5}\,\textrm{d}s$, i.e., $\textrm{d}m \propto s^{-0.5}\,\textrm{d}s$.}
         \label{testdmdr}
   \end{figure}

One well established feature is the waviness of the PSD in the small-grain domain, which is due to the depletion of unbound grains with $\beta>0.5$. This depletion triggers a chain reaction in the PSD: there is an overabundance of grains that should have been destroyed by the absent unbound grains, i.e., those just below the $\beta=0.5$ limit, which in turn causes a depletion of slightly bigger grains that are destroyed by them, causing in turn an overabundance further up the PSD, etc. \citep{camp94,theb03,kriv06}. This waviness is clearly visible in Fig.\,\ref{testdmdr}, displaying the PSD in the birth ring of the simplified inner-parent-body ring case considered in Section \ref{illu}. The first peak in the wavy PSD is located at $\sim 1.5 \, s_\mathrm{cutoff}$ and the first dip around $10\,s_\mathrm{cutoff}$, which is in relatively good agreement with the results obtained by the statistical code of \citet{theb07} (see Fig.\,18 of that paper) for a not-too-dissimilar set-up. The waviness is then progressively damped at larger sizes and is absent in the $s>100\,s_\mathrm{cutoff}$ domain, a result also in good agreement with previous particle-in-a-box studies.

Another important feature of the PSD is that its slope is steeper than -3.5 in the strength regime, i.e., up to objects in the sub-km range. In the $10^{-4}\,$m to $10^{2}\,$m range we find $dm/ds \propto s^{-0.64}$, corresponding to $dN\propto s^{-3.64}\,ds$. This $-3.64$ slope is in remarkable agreement with the -3.7 slope found by \citet{theb07} and the "standard" slope of -3.65 recently found in the extensive study by \citet{gasp12}.

Beyond $s\sim100\,$m, we clearly see another generic feature of debris disc PSDs, namely the flattening of the PSD corresponding to the transition from the tensile-strength to the gravity-dominated regimes \citep[e.g.,][]{lohn08}.

\subsection{Radial Surface Density Profile}

Another well established result is that, beyond a narrow collisionally active ring, the surface density naturally tends towards a radial profile in $r^{-1.5}$. This somehow counter-intuitive feature \footnote{the "intuitive" one being that the profile should be in $\tau \propto r^{-3}$ when (incorrectly) assuming that small grains produced in the ring are simply diluted along their elongated orbits \citep{theb08}} arises because small high-$e$ grains produced in the ring spend most of their time in almost-empty collisionally inactive regions far outside the ring, where they can safely accumulate, while they can only be destroyed during the small fraction of their orbit spent in the production ring \citep{stru06,theb08}. The time $t_\mathrm{ss}$ to reach this radial profile can be relatively long, of the order of the collisional time scale divided by the fraction of time $f$ the smallest bound particles spend in the birth ring:

\begin{equation}
	t_\mathrm{ss} \sim \frac{T_\mathrm{orb}}{2\pi\tau} \,\frac{1}{f},
 	\label{timess}
\end{equation}

where $t_\mathrm{orb}$ is the typical orbital period at the birth ring's location, $\tau$ is the optical depth within the birth ring and $f$ is calculated according to \citet{stru06}'s formula. 

\begin{figure}
   \centering
     \includegraphics[width=9cm]{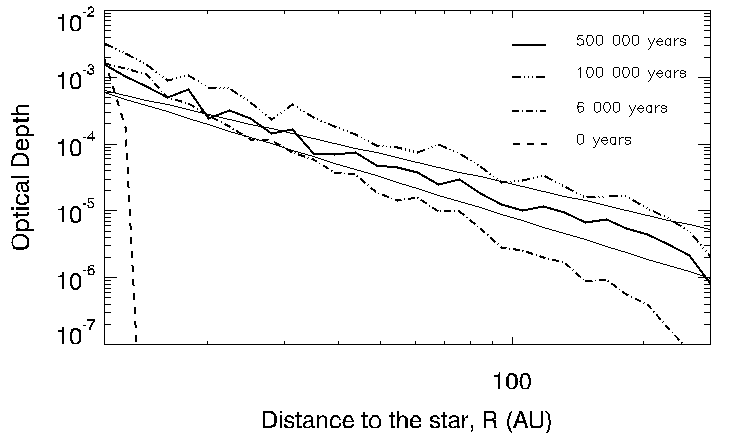}
      \caption{Fiducial inner-parent-body-ring run (see Section \ref{illu}). Azimuthally averaged radial surface density profile at different epochs. The thin black solid lines correspond to profiles in $\tau \propto r^{-1.5}$ and $r^{-2}$ respectively.}
         \label{testprofopt}
   \end{figure}

Fig.~\ref{testprofopt} shows the result found with LIDT-DD simulating the evolution of such an initially confined ring (the set-up considered in section \ref{illu}) over $t=5\times10^{5}\,$years. It can be seen that a steady-state is reached after $\sim10^ {5}\,$years. This value is very close to the theoretical value $t_\mathrm{ss} \sim 1.1\times10^{5}\,$years that can be derived from Equ.~\ref{timess} when taking $t_\mathrm{orb}$ for the middle of the birth ring at 11.5\,AU and calculating $f$ for the $\beta \sim 0.4$ particles that dominate the density profile in the outer regions.
Once steady state is reached, the slope of the radial density profile in the outer regions is close to the theoretical -1.5 value, in accordance with the conclusions of \citet{stru06} and \citet{theb08}. Note, however, that the slope is in fact $\sim -1.7$, i.e., slightly steeper than -1.5. This small difference is expected and is due to the fact that the -1.5 slope is valid for an idealized case where all collisions beyond the ring are neglected, whereas some collisional activity (handled by LIDT-DD) occurs in these outer regions, thus removing some small grains and steepening the profile. After the onset of the quasi steady-state, the shape of the density profile no longer evolves, the only change being the progressive decrease of the total optical depth, which is due to the steady erosion of parent bodies in the birth ring.

\subsection{Dynamically cold system}\label{cold}

Another important result of recent debris disc studies is the peculiar PSD of debris discs that are "dynamically cold", i.e., where the stirring of orbits does not exceed $e \sim 0.01$. For this specific category of discs, \citet{theb08} found that there is a strong depletion of small grains, which arises from an imbalance between the rate at which these grains are \emph{produced} and the rate at which they are \emph{destroyed}. Indeed, while they are mainly produced from collisions amongst larger parent bodies, which will happen at low velocities (and thus produce few small fragments) if $<e>$ is low , they are destroyed by impacts involving themselves, impacts which will always occur at high velocities because the smallest grains are always placed on high-e orbits by radiation pressure regardless of the low dynamical stirring of the system \citep[see discussion in][]{theb08}.

\begin{figure}
   \centering
     \includegraphics[width=9cm]{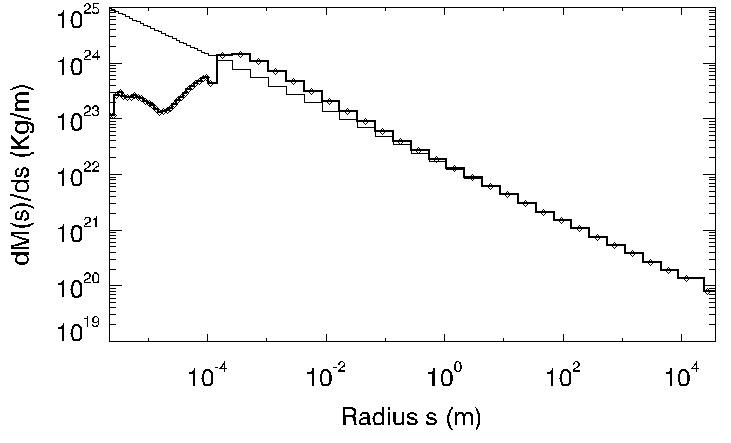}
      \caption{Dynamically cold run (see Section \ref{cold}). Differential particle size distribution, at steady state, expressed in terms of dm/ds, in the main inner ring. The thin black line corresponds to a distribution in $s^{-3.5}\,\textrm{d}s$, i.e., $\textrm{d}m \propto s^{-0.5}\,\textrm{d}s$.}
         \label{testdmdrcold}
   \end{figure}

In Fig.~\ref{testdmdrcold} we display the results obtained with LIDT-DD for such a dynamically cold disc for which $e=0.005$. As expected, the depletion of grains in the $s<0.1$mm range is clearly visible, and the geometrical cross section is dominated by large grains. Note that for excitation values even lower than $e=0.005$, the disc might enter a very different dynamically "very cold" regime for which impacts are no longer fragmenting and mutual accretion is possible.

\begin{figure}
   \centering
     \includegraphics[width=9cm]{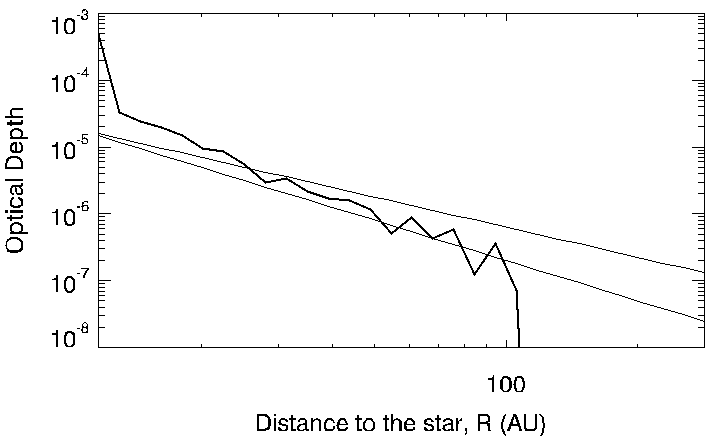}
      \caption{Dynamically cold run (see Section \ref{cold}). Azimuthally averaged radial surface density profile at steady state. The thin black solid lines correspond to radial profiles in $\tau \propto r^{-1.5}$ and $r^{-2}$ respectively.}
         \label{testprofoptcold}
   \end{figure}

Another important characteristics of dynamically cold discs, which directly follows from the dominance of large grains, is that they should display sharp outer edges. This is exactly what we obtain when displaying the system's radial profile in optical depth (Fig.~\ref{testprofoptcold}), which abruptly drops just beyond the production ring. 
The cut-off in the profile further out ($\sim100\,$AU) is due to the limited radial excursion of the smallest particles ($\beta=0.44$) in the simulation, because they are here created from parent bodies on nearly-circular orbits, so that their maximum apoastron will be $\sim a_0/(1-2 \, \beta)=100$ AU, where $a_0=12$ AU is the maximum semi-major axis for a parent body in the main ring.

\section{A case study: break-up of a massive object inside a debris disc}\label{hvc}

The detailed investigation of concrete astrophysical cases exceeds the scope of the present numerical paper and will be the purpose of future studies. However, in order to illustrate the potential of this new code for investigating cases that were so far inaccessible to detailed numerical modelling, we present here preliminary results for the case study of the violent break-up of a large planetesimal in an extended debris disc.
Such violent and transient events have been often invoked to explain the presence of pronounced "clumps" in some resolved debris discs \citep{wyat02,tele05} or the high fractional dust luminosities of some discs, which are either too high or have too rapid variations to be explained with steady collisional cascades \citep{wyat07,liss09,meli12}. Until now, such massive collisions could not be directly simulated. Instead, they were usually modeled with analytical and statistical order-of-magnitude estimates \citep[e.g][]{keny05,liss09}. An alternative approach is the one by \citet{jack12}, using an $N$-body scheme, resolving the initial impact with test particles, each representing a collisional cascade whose mass is steadily decreased using analytical laws.

The most sophisticated attempt at modelling catastrophic collisional events is probably the one by \citet{grig07}. However, this model could not handle the case of a planetesimal exploding inside a disc, and could not be used on long timescales (see Section \ref{spapproach}). The case that was investigated was that of a massive explosion far \emph{outside} (much closer to the central star) the disc, from which only the unbound fragments were followed as they enter an outer disc at high velocities. This set-up allowed to use a bimodal dust population (unbound explosion fragments and larger bound "field" particles) and only required short timescales.

\subsection{Set-up}\label{setupexpl}

As the goal of this example run is not to investigate specific real systems, such as HD172555 \citep{liss09} or TYC 8241 2652 1 \citep{meli12} whose detailed study is deferred to a forthcoming dedicated paper, we consider here only one fiducial case, with one given set of initial parameters, chosen for its simplicity and illustrative virtues \footnote{the detailed study of real astrophysical cases would require the exploration of a vast range of initial parameters and set-ups and is beyond the scope of the present paper}.

We first consider an unperturbed debris disc, similar to the one considered in the test run of Section \ref{illu}, i.e., produced from a parent body ring confined between 11 and 12\,AU. The only difference being a lower optical depth $<\tau>\, =\, 1 \times 10^{-4}$ in order to enhance the contrast between the background debris disc and the fragments created from the impact.

We then consider the catastrophic explosion of a large body releasing a mass $M_\mathrm{Ftot}=2\times 10^{22}$ kg into fragments of sizes $2\,\mu$m\,$< s < 10\,$cm, following a size distribution in $\textrm{d}N \propto s^{-3.5} \textrm{d}s$. This explosion is assumed to take place at $35\,$AU from the star, and we assume that the exploding parent body is on a circular orbit. The mass $M_\mathrm{Ftot}$ corresponds to that of an object of size $\sim 10^{3}\,$km \footnote{In reality, we expect the mass of the shattered planetesimal to exceed $M_\mathrm{Ftot}$, as some mass will be contained in fragments $>10\,$cm, but we do not worry about these issues in this illustrative run and only consider $M_\mathrm{Ftot}$ as a way to scale the magnitude of the initial impact}. We implicitly assume that the collision was just energetic enough so that all of $M_\mathrm{Ftot}$ could escape the potential of the shattered body. To ensure this, we follow the procedure of Jackson \& Wyatt (2012) and release all fragments from a small circle corresponding to 100 physical radiuses of a $\sim 10^{3}\,$ km body with a velocity "kick" randomly distributed between the escape velocity from the shattered body at this distance, $\sim 150$ m.s$^{-1}$, and $1/4$ of the surface escape velocity of this body, $\sim 375$ m.s$^{-1}$.

All relevant initial parameters can be found in Tab.~\ref{tabrelease}.

\begin{table}[h]
\caption{\label{tabrelease}Initial parameters used for the "massive breakup" run. All other parameters are identical to those of the fiducial inner-parent-body ring run described in Tab.~\ref{tabtest}}
\centering
\begin{tabular}{lc}
\hline\hline
Debris ring population\\
\hline
Total mass ($M_\mathrm{tot}$) & $7 \times 10^{23}$ kg \\
Number of initial tracers & 200 000 \\
Optical depth & $<\tau> = 1 \times 10^{-4}$ \\
\hline
Planetesimal release population\\
\hline
Minimum size     & $2.06 \, \mu$m ($\beta$ = 0.44) \\
Maximum size   & 10 cm  \\
Initial size distribution & $\textrm{d}N \propto s^{-3.5} \textrm{d}s$ \\
Initial total mass & $2 \times 10^{22}$ kg \\
Initial velocity of mass released & $V_{kep}$ + $\Delta$ Vlib / 4\\
Number of initial tracers & 10 000 \\
Initial release distance from the star & 35 AU \\
\hline
\end{tabular}
\end{table}

\subsection{Results}

\begin{figure*}

\makebox[\textwidth]{
\includegraphics[scale=0.25]{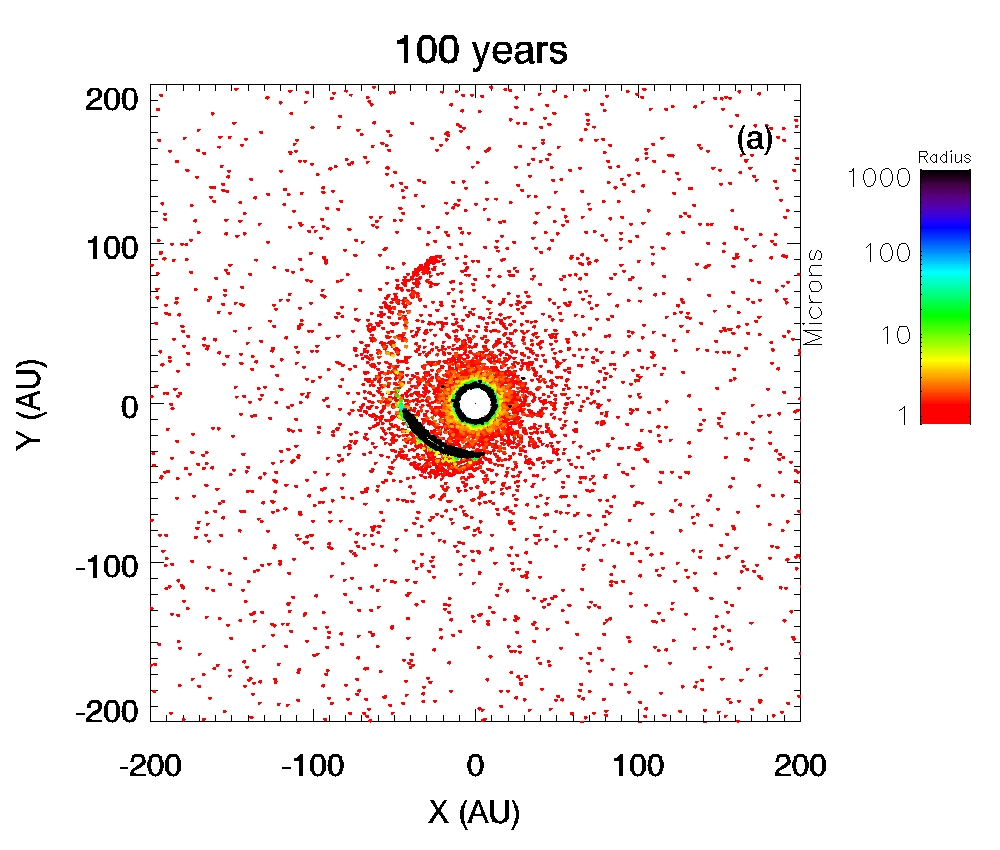}
\includegraphics[scale=0.25]{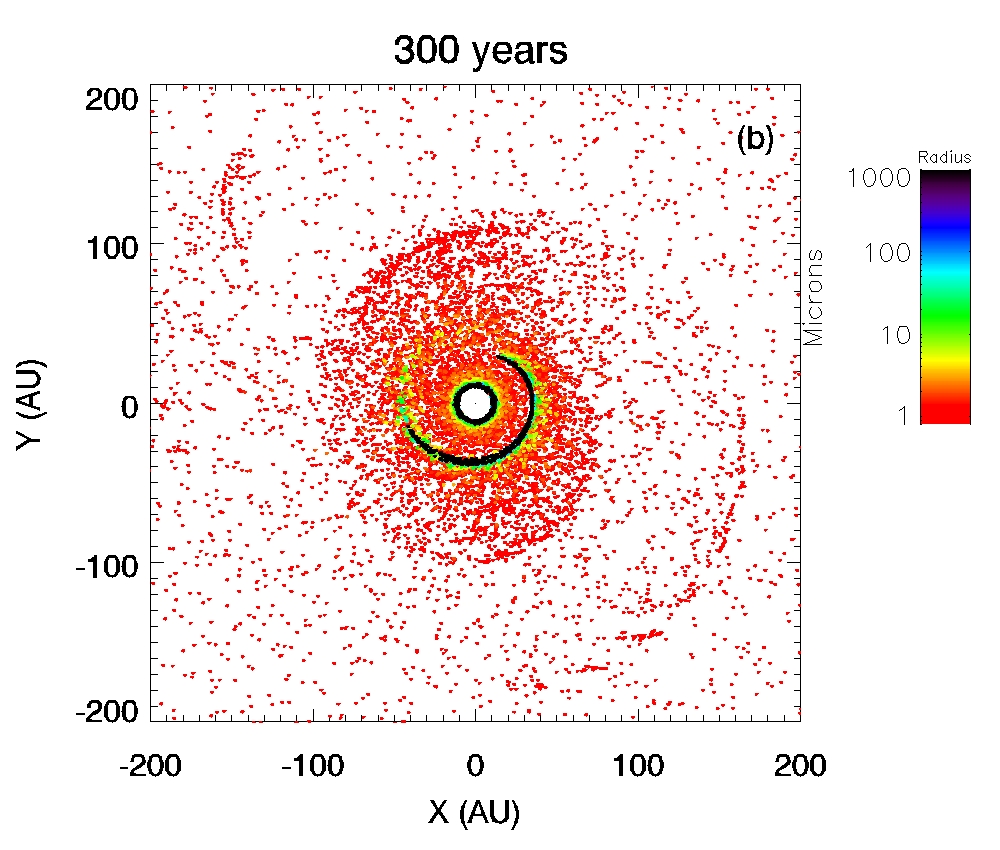}
}
\makebox[\textwidth]{
\includegraphics[scale=0.25]{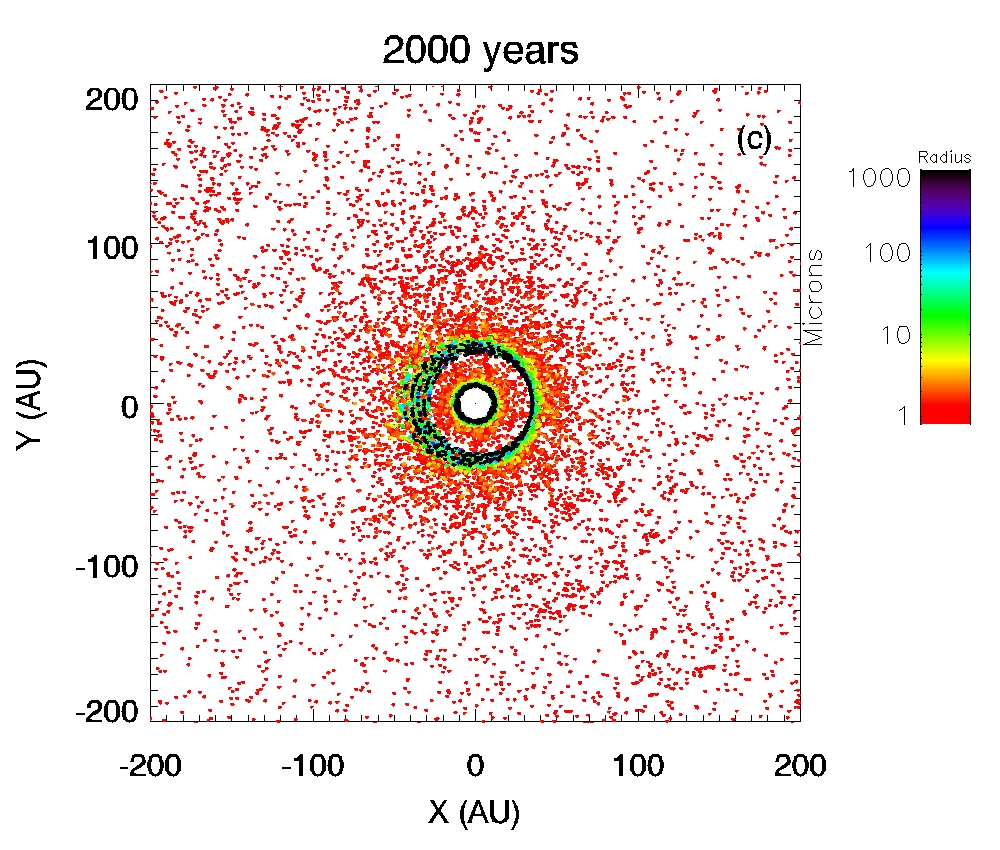}
\includegraphics[scale=0.25]{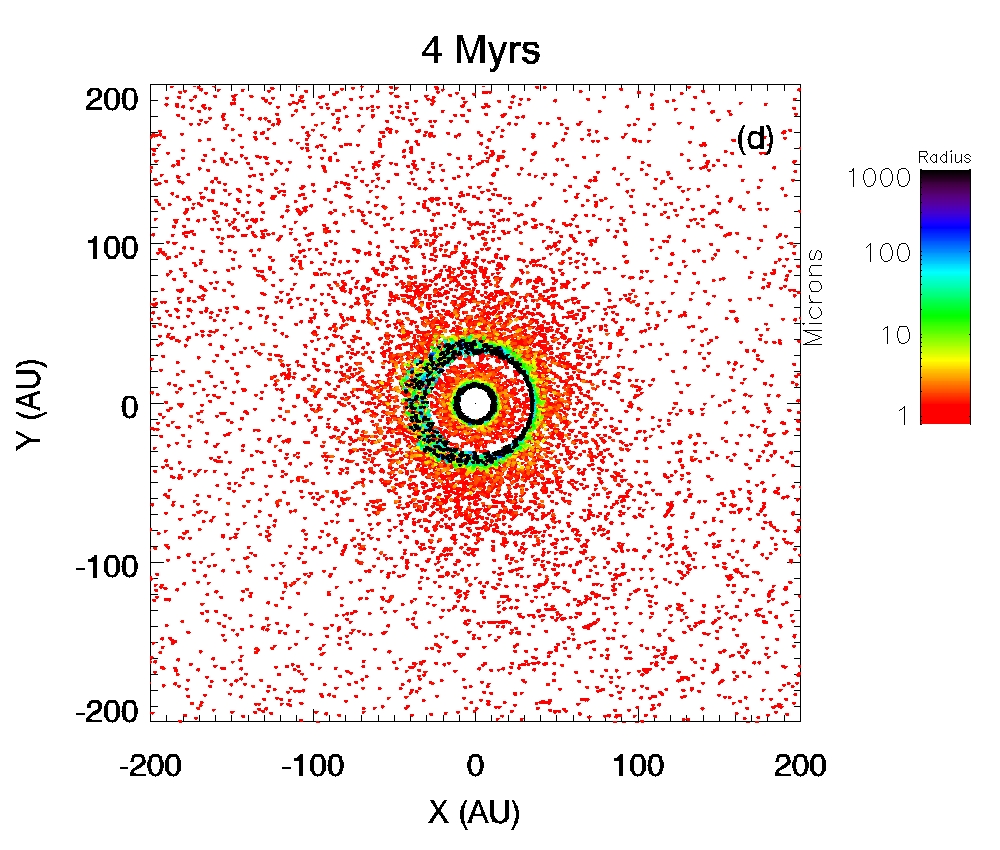}
}
\caption[]{Massive planetesimal breakup run (see Section \ref{setupexpl}). Snapshots of the tracers' positions at 4 different epochs. The colour scale gives the physical sizes, in $\mu$m, of the particles the tracers stand for. The explosion occurs at 35 AU from a $\sim 1000$\,km parent body breakup on a circular orbit (see text for further details).}
\label{crossrun}
\end{figure*}

\begin{figure*}

\makebox[\textwidth]{
\includegraphics[scale=0.2]{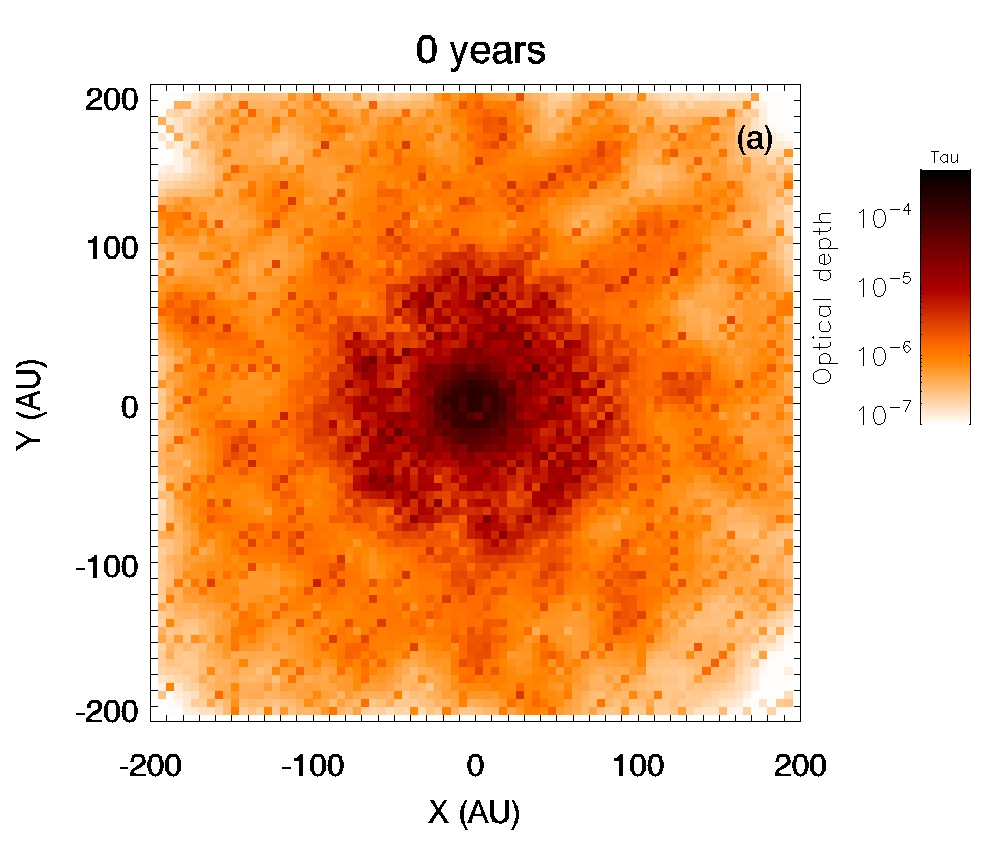}
\includegraphics[scale=0.2]{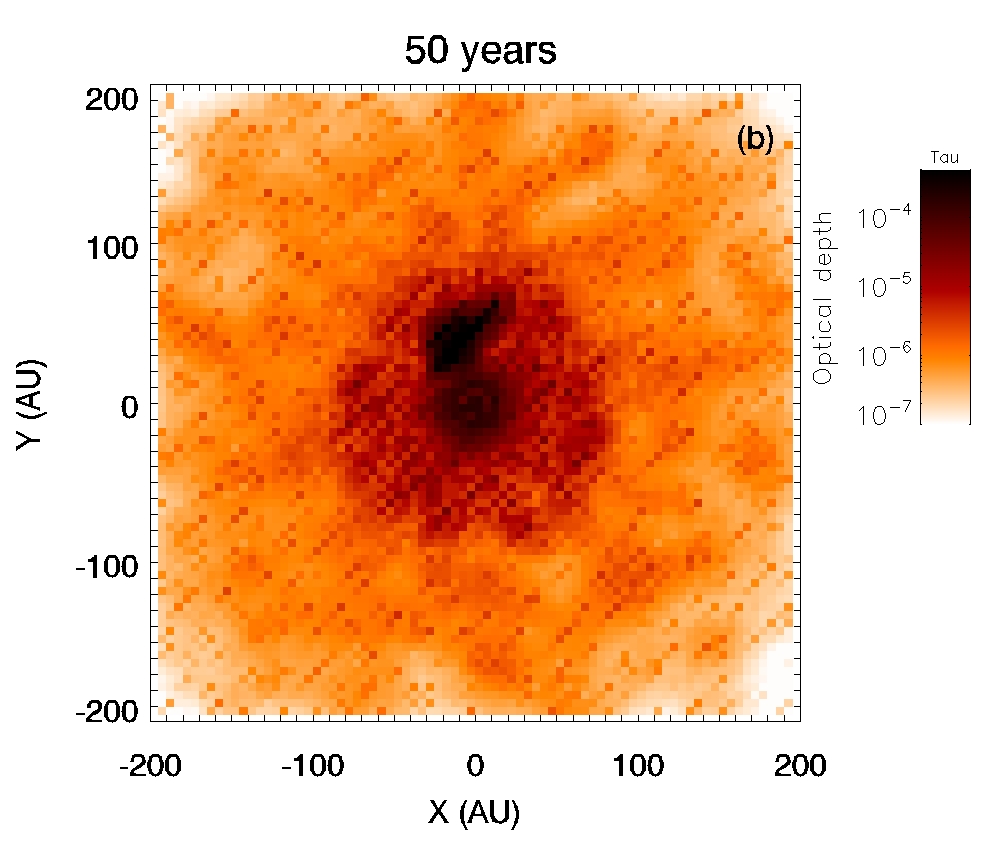}
}
\makebox[\textwidth]{
\includegraphics[scale=0.2]{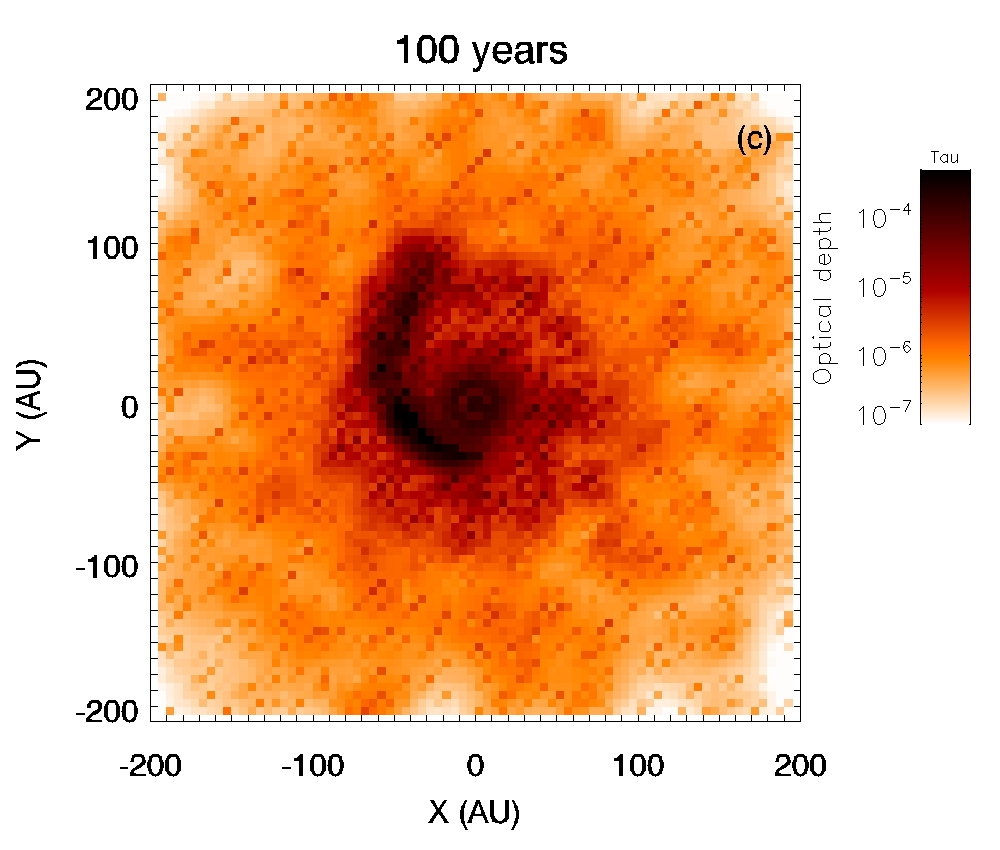}
\includegraphics[scale=0.2]{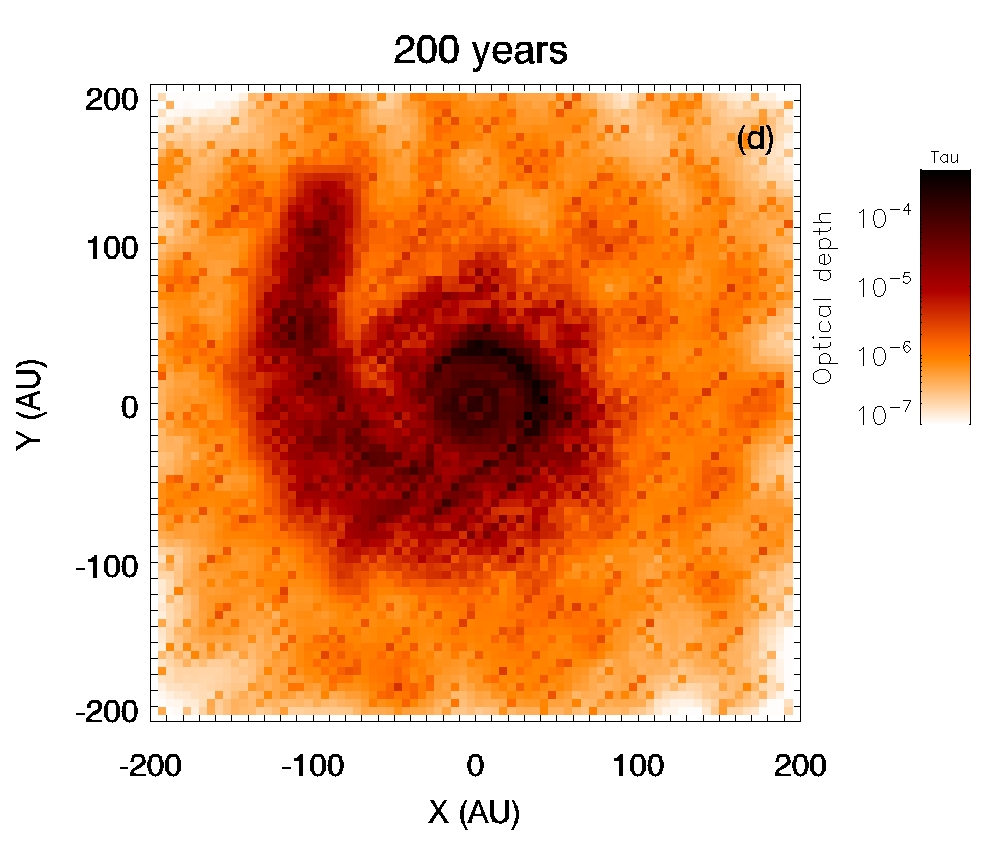}
}
\makebox[\textwidth]{
\includegraphics[scale=0.2]{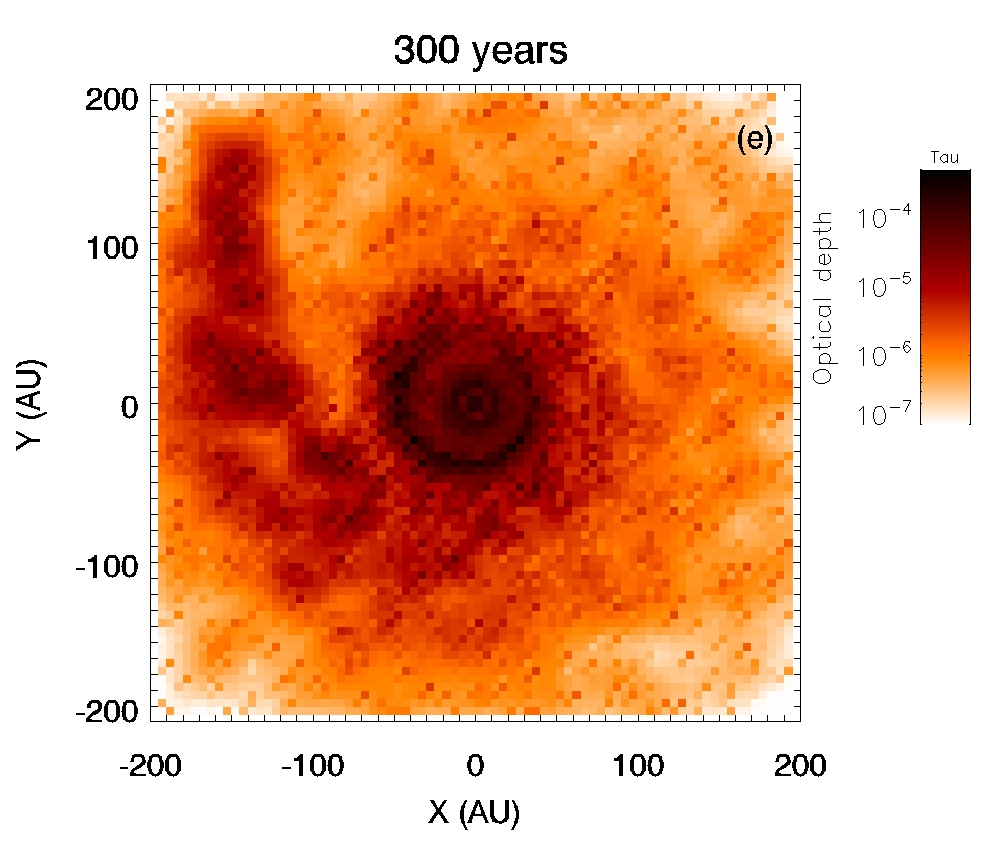}
\includegraphics[scale=0.2]{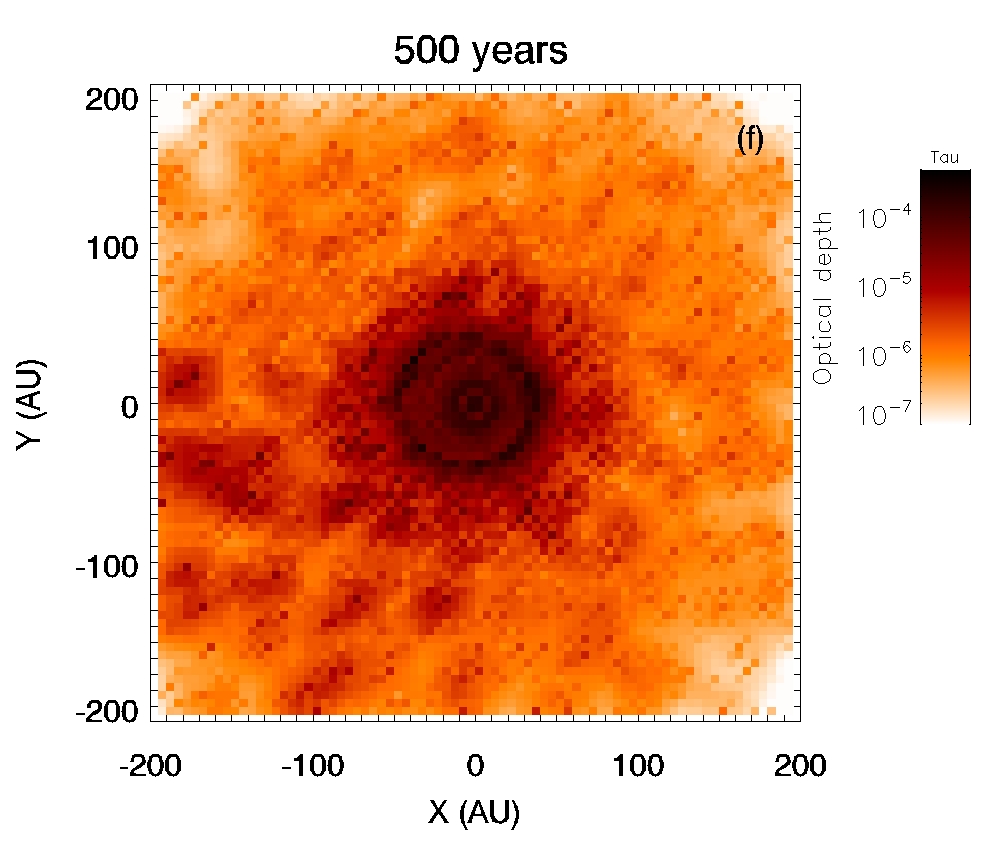}
}
\makebox[\textwidth]{
\includegraphics[scale=0.2]{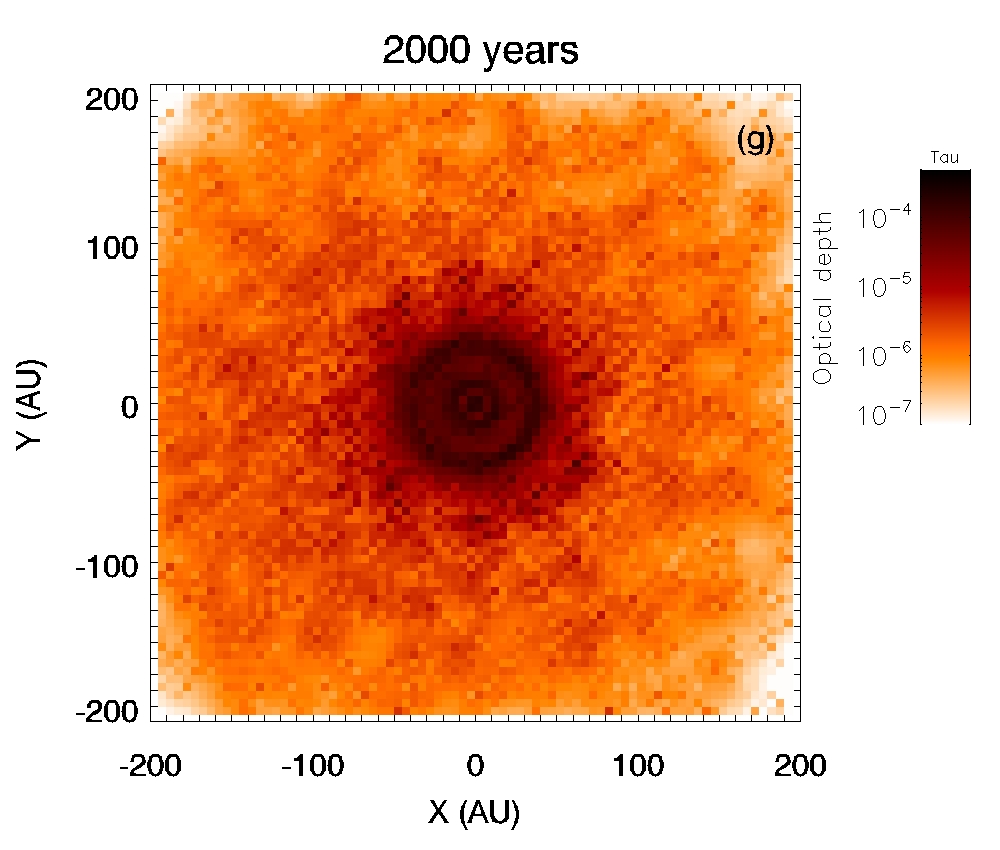}
\includegraphics[scale=0.2]{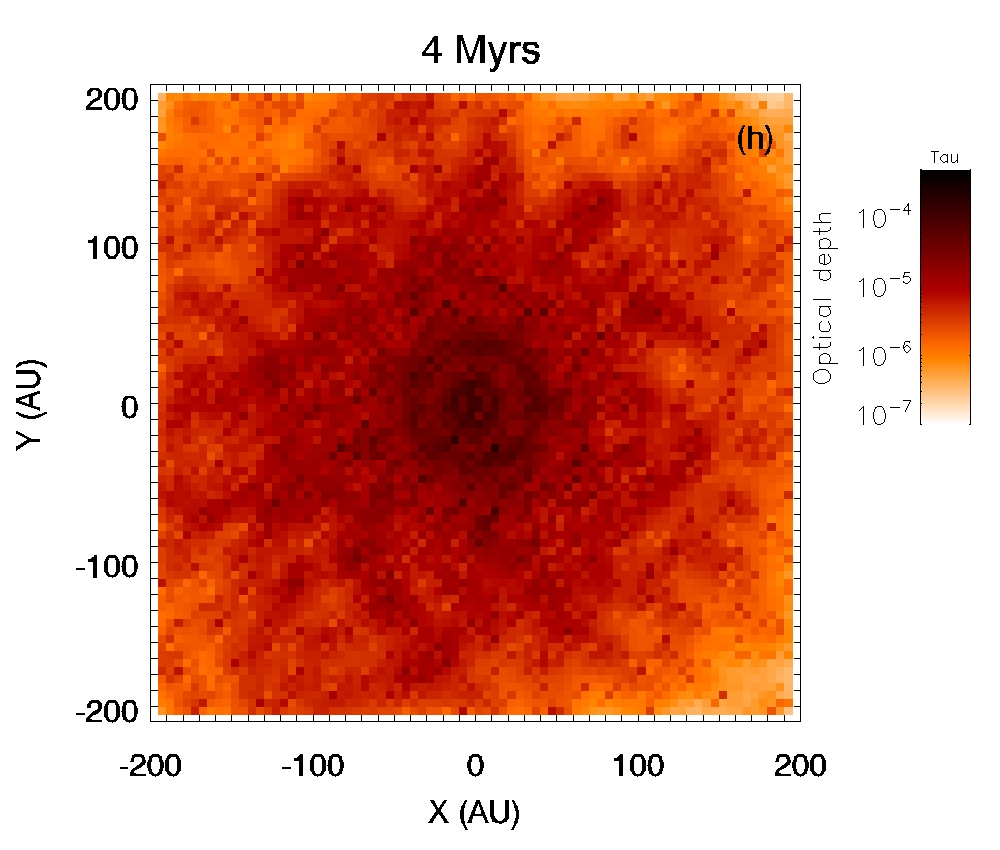}
}
\caption[]{Same run as Fig.~\ref{crossrun}, but this time plotting the total smoothed vertical geometrical optical depth of the system.}

\label{optdepthrun}
\end{figure*}

Fig.~\ref{crossrun} plots, at 4 different epochs, the spatial location of all tracers with a colour scale giving the size bin corresponding to each tracer, while Fig.~\ref{optdepthrun} shows the resulting smoothed 2-D map of the system's geometrical optical depth.

As can be clearly seen, an outward-propagating spiral structure builds up. It corresponds mostly to the smallest explosion fragments, which are pushed on high-eccentricity or unbound orbits by radiation pressure. After $\sim 300\,$years, the smallest grains in the spiral arm have reached the outer limit of the plotted region at $200\,$AU. In parallel, the differential precession of the arm progressively populates the whole $r>35\,$AU region with explosion fragments. The precession rate being faster closer to the star, this populating process proceeds inside out. Meanwhile, at the radial location of the impact around 35AU, the largest fragments progressively form a homogeneous ring because of Keplerian shear.

The development of an outward-propagating spiral after a massive, fragment-releasing impact is a well known feature that has been witnessed in previous $N$-body studies \citep[e.g.,][]{jack12}. However, we are here for the first time able to follow the collisional evolution of the released fragments as they propagate through the disc, as well as the collisional fate of the fragments that stay in the radial region of the explosion. In particular, the code takes into account the production of new generations of collisional fragments from impacts involving the initially released primordial fragments. Such second, third or n$^{th}$ generation debris can significantly contribute to the disc's luminosity in some regions.

LIDT-DD's handling of collisions gives us the possibility to estimate the timescale during which the signature of the explosion remains visible in the disc. We see that the system settles back to an unperturbed 2-D profile in 2 steps. In the outer regions, the propagating spiral and its aftermath are no longer visible, meaning that their signature drops below the noise level in the 2-D map, after $\sim 2000\,$ years (Fig.~\ref{optdepthrun} g). The disappearence of the spiral pattern in $\sim 2000\,$ years is a purely geometrical effect, as this time is approximately the orbital period of the small, high-$\beta$ grains that were released by the initial shattering (this geometrical effect has also been identified by \citet{jack12}). But contrary to the collisionless case of \citet{jack12}, for which, even after the fading of the spiral, the presence of small released grains remained visible in the outer regions for a rather long time, we find here that collisional destruction of these particles prevents them to leave an observable trace once these 2000\,years have elapsed.
The situation is very different at the radial location of the initial release, where the secondary ring made of the largest primordial explosion fragments is still clearly visible. This secondary ring at the explosion's location is much more long lived. It takes another few million years before it is resorbed in the background level (Fig.~\ref{optdepthrun} h). This slow fading of the secondary ring is due to the progressive grinding down of the largest explosion fragments\footnote{The timescale for the secondary ring to totally disappear corresponds to the collisional timescale of the biggest bodies in this ring}, settling into a collisional cascade that produces smaller and smaller debris, which eventually are blown out by radiation pressure.

Note that these two timescales significantly depend on our arbitrary choice of initial parameters, mainly the value for $M_\mathrm{Ftot}$ and the size of the largest explosion fragments, and we insist that this fiducial set-up was not chosen so as to match a specific real astrophysical system. The important point is here that, once a set-up has been chosen, depending on the case that is to be investigated, then LIDT-DD is able to provide a reliable estimate for the survival time of such transient signatures of massive breakups. It is also able to evaluate how the luminosity profiles evolve with time in the wake of the initial violent event.

\subsection{Tracers vs real disc}

The release of the explosion fragments results in a significant increase of the number of tracers in the system (Fig.~\ref{crossrun}). Note that this increase does not reflect the amount of matter that is released in the disc, but the fact that the smallest explosion fragments, which rapidly populate the outer regions because of radiation pressure, have different dynamical characteristics than the matter "at rest" in the unperturbed debris disc. Since the code's structure requires that each collisional "cell" must contain at least 2 tracers for each identified dynamical category (see Section \ref{sorting}), this means that additional tracers will populate the cells. These tracers can either be "primordial" ones released at the moment of the explosion, or can have been created later as the primordial fragments pass through the outer disc and spawn new sub-fragments that need to be represented by new tracers.

As such, the number of tracers that is needed in the system is decorrelated from the amount of mass that is initially released. For example, the number of tracers and their locations would be exactly the same for a simulation with $M_\mathrm{Ftot}/10$ instead of $M_\mathrm{Ftot}$. The difference would be in the number of physical particles each tracer stands for. This is also why the number of tracers does not significantly decrease, even long after the effect of the explosion is no longer visible in the "real" outer disc (see Fig.~\ref{crossrun} c and d), simply because a dynamical category, once created in a cell, is maintained as long as even a little amount of matter corresponds to it. In the present case, as there is always some collisional production of new fragments from the excess matter remaining around 35 AU, there is always some matter that is produced with dynamical characteristics that roughly match those of the first initially released fragments. True, the amount of such newly injected fragments will decrease with time, and they are no longer visible outside of 35 AU on the 2-D maps after $\sim 2000$\,years, but the code is still able to identify and handle them even when they do not show up in the global density maps, hence the still large number of tracers at this stage (see Fig.~\ref{crossrun} d).

\section{Limitations} \label{limit}
 
Despite its many new features and the improvement it brings to disc studies, LIDT-DD, as all numerical codes, has unavoidable limitations. The main ones are related to the complexity of the numerical treatment itself, and especially the tracer-tracer collision treatment procedure (and the tracer creation and reassignment that follows), which is very time-consuming. This limits the number of tracers that can be handled to typically a few $10^{5}$. While this number allows a relatively good spatial sampling, it is at least an order of magnitude less than the number of test particles that can be typically handled in pure $N$-body codes, or more exactly $N-$body codes where the restricted 3-body case applies \citep[e.g.][]{rech09}, so that the spatial resolution will always be less than in such purely dynamical codes. This also limits the timescale that can be investigated with the present LIDT-DD version to a few $10^{6}\,$years. While this is enough to study the evolution of young and bright discs such as $\beta$ Pic or HR4796A, it does not allow to follow the evolution of older systems over their entire lifetime. However, this limitation might not be too constraining, as many processes shaping the structure of even very old discs might develop and evolve on timescales shorter than a few million years. 

The fact that only 2 tracers per dynamical family are kept and that virtual tracers are assigned the same velocity as the larger of the two initial colliding tracers necessarily introduces small errors in the velocity distribution with every time step. These small errors are clearly visible in the form of short-term variations of the system's angular momentum (Fig.~\ref{figangmom}). However, they remain limited in their amplitude (less than $\sim$0.2\%) and have no cumulative effect, since they do not lead to systematic variations of the angular momentum over long timescales.

The limited number of tracers also introduces a relatively high noisiness in the density maps that are obtained. This is due to the finite distances between neighbouring tracers, which can be relatively large in scarcely populated regions. This noisiness can be overcome by averaging surface density over larger regions, but the price to pay is in this case a loss in spatial resolution. Note, however, that this problem affects mainly the regions that are the most dynamically quiet, whereas dynamically perturbed regions, which are usually of main interest in concrete studies, are much more densely populated in terms of tracers, thus allowing a much better spatial resolution (see Fig.~\ref{optdepthrun}).

Another limitation is that a 2-D ($r,\theta$) collisional grid is used to compute size distributions. This implicitly assumes a symmetry in the $Z$ direction for collision rates, an approximation that might be faulty when considering non-planar cases such as an inclined planet. A 3-D collisional grid going in the $Z$ direction as well would solve the problem but would multiply all the calculations by at least a factor $N_z$ ($N_z$ being the number of cells in the $Z$ direction at a fixed $r$ and $\theta$). Note however, that the dynamical evolution of the system is always 3-D.

\section{Summary and perspectives}\label{conclusion}

LIDT-DD is a new code developed to study debris discs, based on a global approach coupling an $N$-body handling of the dynamics to a particle-in-a-box treatment of collisions. It takes its basic architecture from the LIDT3D code developed by \citet{char12}, for the very different context of protoplanetary discs, with "tracers", each representing a whole population of particles of a given size at a given location in the system, whose collisional interactions are handled, with a statistical approach, within superimposed spatial cells. It can, however, be considered as an independent stand-alone code, because of the major modifications and upgrades introduced to account for the very constraining specificities of debris disc physics. The three major such constraints are that 1) impact velocities are much higher in debris discs, so that collisions are highly destructive and fragment-producing; 2) The dynamics is much more complex because, contrary to protoplanetary discs, in the absence of the smoothing effecf of gas drag, initial conditions are not relaxed so that all grains retain the characteristics of where and how they have been produced. As such, there is a very wide range of possible dynamics for same-sized particles located in the same region of the disc: ; and 3) the collisional and dynamical evolution of small grains close to the blow-out limit, those that usually dominate the disc's luminosity, is extremely sensitive to small size variations.

The main specificities of the new code are the following: 
\begin{itemize}
\item All mutual tracer-tracer impacts are handled individually and the produced collisional fragments have orbits derived from those of their collisional progenitors.
\item The collision outcome procedure is now able to handle high-velocity fragmenting impacts and is comparable in sophistication to those implemented in classical particle-in-a-box codes \citep[e.g.,][]{kriv06,theb07}.
\item In addition to sorting tracers by size and spatial location, the code is able to sort and regroup them by dynamical families, using a specially developed hierarchical clustering procedure. This allows LIDT-DD to track down and keep a trace of all the possible dynamical origins for grains present in a given region of the disc.
\item The procedure for tracer creation and reassignment is designed to preserve, in each spatial cell and each size bin, 2 tracers per dynamical family. This prevents both the accidental removal of important dynamical information and the unwanted increase in the number of tracers used by the code.
\end{itemize}

In the absence of any reference result for the coupled dynamical and collisional evolution of discs that could be used as reliable comparison, LIDT-DD has been tested on three simplified cases for which robust results have been obtained in past studies:
\begin{itemize}
\item We are able to retrieve the main features for the particle size distribution of unperturbed collisional debris discs, such as its waviness in the small grain domain and its steeper-than $s^{-3.5}$ slope in the millimetre-to-subkilometre range. The value we find for this slope in our nominal test case is -3.64, very close to the "standard" value of -3.65 inferred by \citet{gasp12}.
\item For the case of an initially narrow ring-like disc, we reproduce the results of \citet{theb08} showing that the surface density profile beyond the initial ring converges towards a radial decrease in $\sim r^{-1.5}$.
\item For the case of a "dynamically cold" disc, with particles on very low orbital eccentricities, we confirm the result obtained by \citet{theb08} and \citet{lohn12} that such a disc is depleted of small grains and dominated by particles bigger than $\sim 10\,s_\mathrm{cutoff}$. 
\end{itemize}
In addition to these comparison tests we also independently check the reliability of our tracer creation and reassignment procedure by verifying that the mass fluxes of outgoing and ingoing particles in a collisional disc at steady state do almost perfectly balance one-another. Moreover, for simplified unperturbed cases, we verify that the angular momentum of the disc is conserved and that collisional energy dissipation leads to the expected decrease of the rms eccentricity of tracers as well as to the expected viscous spread of the disc.

As an illustration of the code's potential to tackle yet-numerically-unexplored astrophysical cases, we consider the case study of the break-up of a massive planetesimal in an extended disc. LIDT-DD is able to follow the collisional fate of the smallest grains released by the break-up, and pushed by radiation pressure on high-$e$ or unbound orbits, as they propagate outward in the disc. We see the expected formation of outward propagating spiral arms, which are resorbed, by collisions and dynamical dilution, after a few dynamical periods. The code is also able to monitor the fate of the large amount of mass that stays at the location of the initial break-up. This matter rapidly forms a ring-like structure, which is much more long-lived than the outbound spiral and gets resorbed into the disc's background after a few million years. It is the first time that a code is able to estimate the amplitude and survival time of the signatures such events can leave in a disc. 

An exhaustive investigation of this crucial issue of massive transient collisional events, exploring the wide range of possible set-ups (location of the break-up, released mass, disc density, etc.), exceeds the scope of the present code-introducing paper. This problem will be addressed in a forthcoming study, whose main objectives will be to explore individual systems with "abnormal" flux excesses, such as HD172555 \citep{liss09,john12} or TYC 8241 2652 1 \citep{meli12}, and also to investigate how generic the massive-break-up scenario can be for explaining \emph{all} bright debris discs with luminosities that cannot be explained by classical collisional cascades.

Another area of interest for LIDT-DD are the planet-disc interactions and the extent to which planetary companions can sculpt debris discs. This issue has been recently explored by both the CGA and DyCoSS algorithms \citep[e.g.][]{kuch10,theb12b}. Although these studies have given promising results, these codes' structures limit them to relatively restrictive set-ups, i.e., systems at steady state, with no more than one planet and with collisional prescriptions assuming that all impacts are fully destructive, thus neglecting all second-generation fragments. LIDT-DD is not bound by these limitations and will thus enable the exploration of transient events, multi-planet systems as well as the feedback of the planetary perturbations on the collisional evolution. We stress that the CGA and DyCoSS codes are by no means obsolete, as they enable a very high spatial resolution that cannot be attained, for the time being, with the present code (see Sec.~\ref{limit}). 

Another potential application of the code is the puzzling case of bright exozodiacal discs, which have been identified by interferometry very close to several main sequence stars such as Vega or Fomalhaut \citep[e.g.][]{absi09,defr11}. The estimated dustiness of these hot exozodis is in most cases far too high to be explained by steady collisional cascade and alternative scenarios have to be considered \citep{menn13}. One possible scenario could be the injection of material from a planetesimal belt, further out in the disc, which is dynamically destabilized by a perturbing planet. This scenario has been proven to be dynamically viable, at least for young systems, by the $N$-body investigation of \citet{bons12}. However, it remains to be seen how this inward scattering of large planetesimals can translate into observable small dust. This crucial issue directly depends on the collisional fate of these bodies and can be investigated by LIDT-DD. Alternative scenarios for explaining exozodis, such as transient massive impacts, falling evaporating bodies \citep{theb01,beus07} or pile up due to the complex interplay of PR drag, sublimation and collisions \citep{koba11}, can in principle also be investigated with LIDT-DD.
More generally, LIDT-DD enables the handling of all complex scenarios where both dynamics and collisions are expected to play an important role in a debris disc evolution.

\begin{acknowledgements}
We thank Jean-Charles Augereau and Amy Bonsor for fruitful discussions. We thank the anonymous referee for helping
improving the quality of the paper. Q.K. acknowledges financial support from the French National Research Agency (ANR) through contract ANR-2010 BLAN-0505-01 (EXOZODI).\end{acknowledgements}

{}
\clearpage


\begin{thebibliography}{}
%
\bibitem[Absil et al.(2009)]{absi09} Absil, Olivier; Mennesson, Bertrand; Le Bouquin, Jean-Baptiste; Di Folco, Emmanuel; Kervella, Pierre; Augereau, Jean-Charles, 2009, ApJ, 704, 150
%
\bibitem[Acke et al.(2012)]{acke12} Acke, B., Min, M., Dominik, C., et al.\ 2012, \aap, 540, A125 
%
\bibitem[Ardila et al.(2005)]{ardi05} Ardila, D.R., Lubow., S.H., Golimovski, D.A., et al., 2005, ApJ, 627, 986
%
\bibitem[Artymowicz \& Clampin(1997)]{arty97} Artymowicz, P., \& Clampin, M.\ 1997, \apj, 490, 863 
%
\bibitem[Augereau et al.(2001)]{auge01} Augereau, J. C.; Nelson, R. P.; Lagrange, A. M.; Papaloizou, J. C. B.; Mouillet, D. 2001, A\&A, 370, 447
%
\bibitem[Augereau \& Papaloizou(2004)]{auge04} Augereau, J.-C., Papaloizou, J.C.B., 2004, A\&A, 414, 1153
%
\bibitem[Bader \& Deuflhard(1983)]{bade83} Bader, G., Deuflhard, P., Numer. Math., 373-398
%
\bibitem[Beauge \& Aarseth(1990)]{beau90} Beauge, C., Aarseth, S.J., 1990, MNRAS, 245, 30
%
\bibitem[Benz \& Asphaug(1999)]{benz99} Benz, W., \& Asphaug, E.\ 1999, \icarus, 142, 5 
%
\bibitem[Beust \& Valiron(2007)]{beus07} Beust, H., \& Valiron, P.\ 2007, \aap, 466, 201
%
\bibitem[Besla \& Wu(2007)]{besl07} Besla, G., \& Wu, Y.\ 2007, \apj, 655, 528
%
\bibitem[Bonsor et al.(2012)]{bons12} Bonsor, A.; Augereau, J.-C.; Thébault, P., 2012, A\&A, 548, 104
%
\bibitem[Booth et al.(2009)]{boot09} Booth, M., Wyatt, M.C., Morbidelli, A., Moro-Mart\'in, A., Levison, H.F., 2009, MNRAS, 399, 385
%
\bibitem[Booth et al.(2013)]{boot13} Booth, M., Kennedy, G., Sibthorpe, B., et al.\ 2013, \mnras, 428, 1263 
%
\bibitem[Campo Bagatin et al.(1994)]{camp94} Campo Bagatin, A., Cellino, A.; Davis, D. R., Farinella, P., Paolicchi, P., 1994, P\&SS, 42, 1079
%
\bibitem[Charnoz et al.(2001)]{char01} Charnoz, S., Th{\'e}bault, P., \& Brahic, A.\ 2001, \aap, 373, 683 
%
\bibitem[Charnoz \& Taillifet(2012)]{char12} Charnoz, S., Taillifet, E., 2012, ApJ, 753, 119
%
\bibitem[Chiang et al.(2009)]{chia09} Chiang, E., Kite, E., Kalas, P., Graham, J.~R., \& Clampin, M.\ 2009, \apj, 693, 734 
%
\bibitem[Debes et al.(2009)]{debe09} Debes, J.~H., Weinberger, A.~J., \& Kuchner, M.~J.\ 2009, \apj, 702, 318 
%
\bibitem[Defr\`ere et al.(2011)]{defr11} Defr\`re, D.; Absil, O.; Augereau, J.-C.; di Folco, E.; Berger, J.-P.; Coudé Du Foresto, V.; Kervella, P.; Le Bouquin, J.-B.; Lebreton, J.; Millan-Gabet, R.; and 3 coauthors, 2011, A\&A, 534, 5
%
\bibitem[Dohnanyi(1969)]{dohn69} Dohnanyi J.S., 1969, JGR 74, 2531
%
\bibitem[Dominik \& Decin(2003)]{domi03} Dominik, C., Decin, G., 2003, ApJ, 598, 626
%
\bibitem[Donaldson et al.(2012)]{dona12} Donaldson, J.~K., Roberge, A., Chen, C.~H., et al.\ 2012, \apj, 753, 147 
%
\bibitem[Ertel et al.(2012)]{erte12b} Ertel, S., Wolf, S., Marshall, J.~P., et al.\ 2012, \aap, 541, A148 
%
\bibitem[Fujiwara et al.(1977)]{fuji77} Fujiwara, A., Kamimoto, G., \& Tsukamoto, A.\ 1977, \icarus, 31, 277  
%
\bibitem[G\'asp\'ar et al.(2012)]{gasp12} G\'asp\'ar, A., Psaltis, D., Rieke, G.~H., Ozel, F., 2012, ApJ, 754, 74
%
\bibitem[G{\'a}sp{\'a}r et al.(2013)]{gasp13} G{\'a}sp{\'a}r, A., Rieke, G.~H., \& Balog, Z.\ 2013, \apj, 768, 25 
%
\bibitem[Golimovski et al.(2006)]{goli06} Golimowski, D. A., et al. 2006, ApJ, 131, 3109
%
\bibitem[Grigorieva et al.(2007)]{grig07} Grigorieva, A., Artymowicz, P., Thebault, P., 2007, A\&A, 461, 537
%
\bibitem[Housen \& Holsapple(1990)]{hous90} Housen, K.~R., \& Holsapple, K.~A.\ 1990, \icarus, 84, 226 
%
\bibitem[Jackson \& Wyatt(2012)]{jack12} Jackson, A.~P., \& Wyatt, M.~C.\ 2012, \mnras, 425, 657 
%
\bibitem[Johnson et al.(2012)]{john12} Johnson, B. C.; Lisse, C. M.; Chen, C. H.; Melosh, H. J.; Wyatt, M. C.; Thebault, P.; Henning, W. G.; Gaidos, E.; Elkins-Tanton, L. T.; Bridges, J. C.; Morlok, A., 2012, ApJ, 761, 45
%
\bibitem[Kalas et al.(2005)]{kala05} Kalas, P., Graham, J. R., \& Clampin, M. 2005, Nature, 435, 1067
%
\bibitem[Kenyon \& Bromley(2002)]{keny02} Kenyon, S.~J., \& Bromley, B.~C.\ 2002, \aj, 123, 1757 
%
\bibitem[Kenyon \& Bromley(2005)]{keny05} Kenyon, S.J., Bromley, B.C., 2005, AJ, 130, 269
%
\bibitem[Kobayashi et al.(2011)]{koba11} Kobayashi, H., Kimura, H., Watanabe, S.-i., Yamamoto, T., 2011, Earth, Planets, and Space, 63, 1067 
%
\bibitem[Krivov et al.(2006)]{kriv06} Krivov, A.~V., L{\"o}hne, T., \& Srem{\v c}evi{\'c}, M.\ 2006, \aap, 455, 509 
%
\bibitem[Krivov(2010)]{kriv10} Krivov, A.~V.\ 2010, Research in Astronomy and Astrophysics, 10, 383 
%
\bibitem[Kuchner \& Holman(2003)]{kuch03} Kuchner, M.J., Holman, M.J., 2003, ApJ, 588, 1110
%
\bibitem[Kuchner \& Stark(2010)]{kuch10} Kuchner, M.~J., \& Stark, C.~C.\ 2010, \aj, 140, 1007 
%
\bibitem[Lestrade et al.(2012)]{lest12} Lestrade, J.-F., Matthews, B.~C., Sibthorpe, B., et al.\ 2012, \aap, 548, A86 
%
\bibitem[Levison et al.(2012)]{levi12} Levison, H.~F., Duncan, M.~J., \& Thommes, E.\ 2012, \aj, 144, 119 
%
\bibitem[Lynden-Bell \& Pringle(1974)]{lynd74} Lynden-Bell, D., \& Pringle, J.~E.\ 1974, \mnras, 168, 603 
%
\bibitem[Lisse et al.(2009)]{liss09} Lisse, C.~M., Chen, C.~H., Wyatt, M.~C., et al.\ 2009, \apj, 701, 2019 
%
\bibitem[Lithwick \& Chiang(2007)]{lith07} Lithwick, Y., \& Chiang, E.\ 2007, \apj, 656, 524
%
\bibitem[L{\"o}hne et al.(2008)]{lohn08} L{\"o}hne, T., Krivov, A.~V., \& Rodmann, J.\ 2008, \apj, 673, 1123 
%
\bibitem[L{\"o}hne et al.(2012)]{lohn12} L{\"o}hne, T., Augereau, J.-C., Ertel, S., et al.\ 2012, \aap, 537, A110 
%
\bibitem[Marzari \& Scholl(2000)]{marz00} Marzari, F., \& Scholl, H.\ 2000, ApJ, 543, 328 
%
\bibitem[Marzari \& Thebault(2011)]{marz11} Marzari, F., \& Th{\'e}bault, P.\ 2011, \mnras, 416, 1890 
%
\bibitem[Melis et al.(2012)]{meli12} Melis, C., Zuckerman, B., Rhee, J.~H., et al.\ 2012, \nat, 487, 74
%
\bibitem[Mennesson et al.(2013)]{menn13} Mennesson, B.; Absil, O.; Lebreton, J.; Augereau, J.-C.; Serabyn, E.; Colavita, M. M.; Millan-Gabet, R.; Liu, W.; Hinz, P.; Thebault, P., 2013, ApJ, 643, 119
%
\bibitem[Mouillet et al.(1997)]{moui97} Mouillet, D.; Larwood, J. D.; Papaloizou, J. C. B.; Lagrange, A. M., 1997, MNRAS, 292, 896
%
\bibitem[Raymond et al.(2011)]{raym11} Raymond, S., Armitage, P.J.; Moro-Mart\'in, A.; Booth, M.; Wyatt, M.C.; Armstrong, J. C.; Mandell, A.M.; Selsis, F.; West, A.A., 2011, A\&A, 530, 62
%
\bibitem[Reche et al.(2009)]{rech09} Reche, R.; Beust, H.; Augereau, J.-C., 2009, A\&A, 463, 661
%
\bibitem[Robertson(1937)]{robe37} Robertson, H.~P.\ 1937, \mnras, 97, 423 
%
\bibitem[Stark \& Kuchner(2009)]{star09} Stark, C. C., Kuchner, M.J., 2009, ApJ, 707, 543
%
\bibitem[Strubbe \& Chiang(2006)]{stru06} Strubbe, L.E., Chiang, E.I., 2006, ApJ, 648, 652
%
\bibitem[Takeuchi \& Artymowicz(2001)]{take01} Takeuchi, T., Artymowicz, P., 2001, ApJ, 557, 990
%
\bibitem[Telesco et al.(2005)]{tele05} Telesco, C.~M., Fisher, R.~S., Wyatt, M.~C., et al.\ 2005, \nat, 433, 133 
%
\bibitem[Thebault \& Brahic(1998)]{theb98} Thebault, P., \& Brahic, A.\ 1998, \planss, 47, 233 
%
\bibitem[Thebault(2009)]{theb09} Th{\'e}bault, P.\ 2009, \aap, 505, 1269 
%
\bibitem[Thebault(2012)]{theb12a} Thebault, P., 2012, A\&A, 537, 65
%
\bibitem[Thebault \& Augereau (2007)]{theb07} Thebault, P., Augereau, J. C.,  2007, A\&A, 472, 169
%
\bibitem[Thebault et al.(2003)]{theb03} Thebault, P., Augereau, J.~C., \& Beust, H.\ 2003, \aap, 408, 775 
%
\bibitem[Thebault \& Beust (2001)]{theb01} Thebault, P., Beust, H.,  2001, A\&A, 376, 621
%
\bibitem[Thebault et al.(2012)]{theb12b} Thebault, P., Kral, Q., \& Ertel, S.\ 2012, \aap, 547, A92 
%
\bibitem[Thebault et al.(2010)]{theb10} Thebault, P., Marzari, F., Augereau, J.-C., 2010, A\&A, 524, 13
%
\bibitem[Thebault \& Wu(2008)]{theb08} Thebault, P., Wu, Y., 2008, A\&A, 481, 713
%
\bibitem[Ward(1963)]{ward63} Ward, J. H., Jr. ,1963, Hierarchical Grouping to Optimize an Objective Function, Journal of the American Statistical Association, 48, 236Ð244
%
\bibitem[Wetherill \& Stewart(1993)]{weth93} Wetherill, G.~W., \& Stewart, G.~R.\ 1993, \icarus, 106, 190 
%
\bibitem[Wyatt et al.(1999)]{wyat99} Wyatt, M.~C., Dermott, S.~F., Telesco, C.~M., et al.\ 1999, \apj, 527, 918 
%
\bibitem[Wyatt \& Dent(2002)]{wyat02} Wyatt, M.~C., \& Dent, W.~R.~F.\ 2002, \mnras, 334, 589 
%
\bibitem[Wyatt et al.(2007)]{wyat07} Wyatt, M.~C., Smith, R., Greaves, J.~S., et al.\ 2007, \apj, 658, 569 
%
\bibitem[Wyatt(2008)]{wyat08} Wyatt, M.~C.\ 2008, \araa, 46, 339 
%
\bibitem[Wyatt et al.(2011)]{wyat11} Wyatt, M.~C., Clarke, C.J., Booth, M., 2011, CeMDA, 111, 1  
%
\bibitem[Wyatt et al.(2012)]{wyat12} Wyatt, M.~C., Kennedy, G., Sibthorpe, B., et al.\ 2012, \mnras, 424, 1206
%
\end{thebibliography}
\end{document}